\documentclass[preprint]{elsarticle}
\journal{Computer and Fluids}

\newcommand{\bsub}{\begin{subequations}}
\newcommand{\esub}{\end{subequations}}

\newcommand{\ka}{\kappa}

\newcommand{\po}{\mbox{\boldmath $\omega$}}

\newcommand{\pt}{\mbox{\boldmath $\tau$}}

\newcommand{\bS}{\mathbf  S}

\newcommand{\pU}{\textbf{\emph{U}}}

\newcommand{\pe}{\textbf{\emph{e}}}
\newcommand{\pf}{\textbf{\emph{f}}}

\newcommand{\pn}{\textbf{\emph{n}}}

\newcommand{\pu}{\textbf{\emph{u}}}

\newcommand{\px}{\textbf{\emph{x}}}

\newcommand{\pat}{\partial}
\newcommand{\na}{\nabla}
\newcommand{\x}{\times}

\newcommand{\beq}{\begin{equation}}
\newcommand{\eeq}{\end{equation}}
\newcommand{\bsubeq}{\begin{subequations}}
\newcommand{\esubeq}{\end{subequations}}
\newcommand{\beqn}{\begin{eqnarray}}
\newcommand{\eeqn}{\end{eqnarray}}
\newcommand{\fr}{\frac}
\newcommand{\lb}{\label}
\newcommand{\er}{\eqref}

\usepackage{hyperref}
\hypersetup{colorlinks,allcolors=blue}
\usepackage{amsmath}
\usepackage{multirow}
\usepackage{amsfonts,amssymb}
\usepackage{graphicx}
\usepackage{subfigure}
\usepackage[rel]{overpic}

\begin{document}

\begin{frontmatter}

\title{\textbf{Wall modeled immersed boundary method for high Reynolds number flow over complex terrain}}

\author{Luoqin Liu\corref{mycorrespondingauthor}}
\cortext[mycorrespondingauthor]{Corresponding author: Physics of Fluids Group, Max Planck Center Twente for Complex Fluid Dynamics, University of Twente, 7500 AE Enschede, The Netherlands}
\ead{luoqin.liu@utwente.nl}

\author{Richard J. A. M. Stevens}

\address{Physics of Fluids Group, Max Planck Center Twente for Complex Fluid Dynamics, University of Twente, 7500 AE Enschede, The Netherlands}

\begin{abstract}
Modeling the effect of complex terrain on high Reynolds number flows is important to improve our understanding of flow dynamics in wind farms and the dispersion of pollen and pollutants in hilly or mountainous terrain as well as the flow in urban areas. Unfortunately, simulating high Reynolds number flows over complex terrain is still a big challenge. Therefore, we present a simplified version of the wall modeled immersed boundary method by Chester et al. (\textit{J. Comput. Phys.} 2007; \textbf{225}: 427-448). By preventing the extrapolation and iteration steps in the original method, the proposed approach is much easier to implement and more computationally efficient. Furthermore, the proposed method only requires information that is available to each processor and thus is much more efficient for simulations performed on a large number of cores. These are crucial considerations for algorithms that are deployed on modern supercomputers and will allow much higher grid resolutions to be considered. 
We validate our method against wind tunnel measurements for turbulent flows over wall-mounted cubes, a two-dimensional ridge, and a three-dimensional hill. We find very good agreement between the simulation results and the measurement data, which shows this method is suitable to model high Reynolds number flows over complex terrain.
\end{abstract}


\end{frontmatter}

\section{Introduction}

Wind power, as a clean and renewable energy source, has developed rapidly over recent decades. So far, most wind farm studies have focused on idealized terrain in which a uniform surface roughness height is used to model ground effects \citep{moe84, alb99, bou05, cal10, ste17}. However, there are many viable wind farms locations in mountainous or hilly areas \citep{pol12, con16, alf17,liu20b}, or in the vicinity of large urban centres. Complex terrain topographies strongly influence flow characteristics such as wind speed and direction and turbulence intensity. All these flow properties are important factors for wind energy assessment, short-term wind forecasting, and optimal wind farm sitting. Therefore, it is important to predict wind characteristics over complex terrain accurately.

To study the flow dynamics in wind farms situated in complex terrain in detail, we need a state-of-the-art simulation method that can capture the dynamics in atmospheric boundary layers over complex terrain accurately. Large-eddy simulations have become a valuable research tool to simulate turbulent flow in wind farms. The reason is that large-eddy simulations can capture temporal fluctuations well, while this method is less computationally expensive than direct numerical simulations \citep{wu12, ste14b, nil15, liu19c, mar19}. To reliably and efficiently handle complex terrains, the immersed boundary method has become popular \citep{iac03, mit05, che07}.

Previously, \citet{che07} developed an immersed boundary method for large eddy simulations of high Reynolds number flows. The three main steps in their algorithm are: (1) a near-wall model to determine the stresses in the fluid band region; (2) a linear extrapolation to obtain the stresses in a solid band region around the body; and (3) a successive over-relaxation iteration method to smooth the stress field inside the body. The accuracy of this method has been evaluated by the agreements with experimental data \citep{gra12, die13}. This method, however, is not computationally efficient as it requires the exchange of a significant amount of information between neighbouring processors. Given the trend that simulations are performed using an ever-increasing number of cores, we believe that it is essential to develop a method that is very efficient in highly parallel computations. 

In this paper, we present and validate a simplified version of the wall modeled immersed boundary method by \citet{che07} that can be used for simulations of high Reynolds number flows over complex terrain. We achieve this by ensuring that the proposed method is easy to implement and only uses information that is available to each processor to limit communication between processors. 
In Section~\ref{sec.LES} we present the governing equations and numerical implementation of the large-eddy simulation method in which the wall modeled immersed boundary method is implemented. The details of the immersed boundary approach are presented in Section~\ref{sec.immersed boundary method}. Section~\ref{sec.applications} discusses the validation and comparisons to wind tunnel measurement data. Finally, Section~\ref{sec.conclusion} summarizes the conclusions.

\section{Large-eddy simulation}
\lb{sec.LES}

Large-eddy simulations are widely used to study high Reynolds number turbulent flows. In this simulation method, the large scale flow features are explicitly resolved, while small scale dissipation is parametrized by modeling the sub-grid scale stresses. In this section, we describe the details of the employed large-eddy simulation method.

\subsection{Governing equations}
\label{sec.gov-eqs}

We solve the spatially-filtered unsteady incompressible Navier-Stokes equations. In particular, the continuity and momentum equations are written as
\beqn 
\lb{eq.mass}
\na \cdot \widetilde{\pu} &=& 0, \\
\lb{eq.momentum}
\pat_t \widetilde{\pu} + \widetilde{\po} \times \widetilde{\pu} &=& \pf - \na \widetilde {p} - \na \cdot \pt.
\eeqn
Here, $\widetilde \pu$ is the velocity, $\widetilde{\po} = \na \times \widetilde \pu$ is the vorticity, 
$\widetilde p$ is the modified pressure, $\pt$ is the deviatoric part of the sub-grid scale stress tensor, and $\pf$ is the external force, which can originate from different sources such that 
\beq\lb{eq.force}
\pf = \pf_{\rm pre} + \pf_{\rm Cor} + \pf_{\rm Buoy} + \pf_{\rm turb} + \pf_{\rm IB},
\eeq
where $\pf_{\rm pre}$ is the constant pressure gradient that drives the flow, $\pf_{\rm Cor}$ is the Coriolis force due to the Earth's rotation, $\pf_{\rm Buoy}$ is the buoyancy force due to the temperature gradient, $\pf_{\rm turb}$ is the force due to wind turbines, and $\pf_{\rm IB}$ is the virtual force due to the immersed boundary. In this study $\pf_{\rm Cor}$, $\pf_{\rm Buoy}$ and $\pf_{\rm turb}$ are absent. 

\subsection{Smagorinsky Model}

For closure of Eq.~\er{eq.momentum} it is necessary to obtain an expression for $\pt$ in terms of the resolved fields. For simplicity, the sub-grid scale stress tensor is modeled using the Smagorinsky model \citep{sma63},
\beq \lb {eq.t-ij-1963}
\pt = - 2 \nu_t \widetilde {\bS}, \quad
\widetilde {\bS} \equiv \frac{1}{2} \left[ \na \widetilde {\pu} + (\na \widetilde {\pu})^T \right],
\eeq
where the superscript $T$ denotes the matrix transpose, and $\nu_t$ is the eddy viscosity, which is defined as
\beq \lb {eq.nu-t}
\nu_t = (C_s l)^2 |\widetilde {S}|, \quad 
|\widetilde {S}| \equiv \sqrt{ 2 \widetilde {\bS} : \widetilde {\bS} }.
\eeq
Here $C_s$ is the Smagorinsky coefficient, and $l$ is a representative length scale at which energy passes from the resolved to the sub-grid field. We use a Smagorinsky coefficient $C_{s} = 0.16$. On uniform grids, and far away from the wall, the characteristic length scale $l$ is constant and defined by
\beq \lb {eq.smag}
l =(\Delta x \Delta y \Delta z)^{1/3},
\eeq
where $\Delta x, \Delta y, \Delta z$ are the grid spacings in $x,y,z$ directions, respectively.
However, close to the wall, this length scale can no longer be parametrized by a constant. Instead, it decreases due to the no-slip condition at the wall. For this, we use the phenomenological wall damping function proposed by \citet{mas92}, which states that
\beq \lb {eq.smag-1}
\frac{1}{\lambda^n} = \frac{1}{\lambda_0^n} + \frac{1}{[\kappa (z+z_0)]^n},
\eeq
where $\ka = 0.4$ is the von K\'arm\'an constant, $n=2$ is the damping exponent, $z_0$ is the roughness height, and $\lambda = C_s l$ is the sub-grid scale mixing length with $\lambda_0$ representing the value far away from the wall.

\subsection{Boundary conditions}
\label{sec.les-bc}

We use periodic boundary conditions in the horizontal directions. The top boundary is impenetrable in combination with a no-stress boundary conditions, such that at the top of the domain
\beq \lb {eq.bc-top}
\tau_{xz}=\tau_{yz} = \widetilde{w} = 0.
\eeq
Modeling of the boundary condition at the lower wall is crucial to capture the turbulent structures in the boundary layer correctly. The reason is that turbulence is produced as a result of drag at the wall, which depends on the properties of the surface, and the instantaneous properties of the resolved flow in the immediate vicinity of the wall. In the simulations, the surface shear stress $\tau_w$ is computed at the lower boundary using the local roughness height, $z_0$, and the velocity close to the wall by integrating the logarithmic velocity gradient from $z=z_0$ to $z=z_1$ \citep{moe84, bou05}, such that 
\beq \lb {eq.bc-bottom}
\tau_w = - \left[ \frac{ \kappa U_r }{ \ln (z_1/z_0)} \right]^2, 
\eeq
where $z_1 = \Delta z/2$ is the height of the first node above the wall and $U_r = \sqrt{ \langle \widetilde u \rangle^2 + \langle \widetilde v \rangle^2}$ is the smoothed velocity at $z=z_1$. The smoothing is performed using a spectral cutoff filter with a $2\Delta$ filter width. The stress is distributed in the two horizontal directions as follows
\beq \lb {eq.bc-bottom-1}
\left. \tau_{xz} \right|_{\rm wall}= \tau_w \left. \frac{ \langle \widetilde{u} \rangle }{ U_r }\right|_{z=z_1}, \quad \left. \tau_{yz} \right|_{\rm wall}= \left. \tau_w \frac{ \langle \widetilde{v} \rangle }{ U_r }\right|_{z=z_1}. 
\eeq
The vertical velocity boundary condition at the bottom is 
\beq \lb {eq.bc-bottom-u3}
\widetilde{w} = 0.
\eeq

\subsection{Numerical treatment}\label{sec.les-numer}
We use a uniform mesh in all directions such that the grid spacing in streamwise ($x$), spanwise ($y$), and vertical ($z$) direction are defined as
\beq \lb {eq.dxdydz}
\Delta x = \frac{L_x}{N_x}, \quad
\Delta y = \frac{L_y}{N_y}, \quad
\Delta z = \frac{L_z}{N_z-1}, \quad
\eeq 
respectively. Here $L_x$, $L_y$, and $L_z$ are the lengths of the domain in the respective directions, and $N_x$, $N_y$, and $N_z$ the number of grid points in each direction.

The mesh is staggered in the vertical direction, with the vertical velocity ($w$) stored halfway between the other variables. The horizontal directions are treated with pseudo-spectral differentiation, such that
\beqn \lb {eq.dudx}
\pat_x A(x,y,z) &=& \sum_{k_x, k_y} i k _x \widehat A(k_x, k_y, z) e^{ i(k_x x + k_y y) }, \\
\pat_y A(x,y,z) &=& \sum_{k_x, k_y} i k_y \widehat A(k_x, k_y, z) e^{ i(k_x x + k_y y) }. 
\eeqn
Here $\widehat A$ is the complex Fourier amplitude associated with the physical space variable $A$, and $k_x$ and $k_y$ are the wavenumbers in the $x$ and $y$ directions. The vertical direction is discretised using the 2nd-order central finite difference, 
\beqn \lb {eq.dudz}
\pat_z A_{i,j,k} &=& \frac{A_{i,j,k}-A_{i,j,k-1}}{\Delta z}, \quad 
A = \widetilde u, \widetilde v, \widetilde p; \\
\lb {eq.dwdz}
\pat_z A_{i,j,k} &=& \frac{A_{i,j,k+1}-A_{i,j,k}}{\Delta z}, \quad 
A = \widetilde w. 
\eeqn
As a result the derivatives of the respective variables are stored halfway between the nodes of the original variable. Given that we have a uniform grid this means that the derivatives of quantities that are discretised on the $uv$-nodes are located at the $w$ nodes and vice versa. 

The velocity field is integrated in time using a 2nd-order Adams--Bashforth method \citep{can88}. To ensure the divergence-free condition for the velocity field the projection method is used \citep{cho68}. First, the intermediate velocity $\widetilde \pu^{*}$ is calculated, 
\beq \lb{eq.u-inter}
\frac{ \widetilde \pu^{*} - \widetilde \pu^t }{ \Delta t } = \frac{3}{2} \textrm{RHS}'^t - \frac{1}{2} \textrm{RHS}^{t-\Delta t},
\eeq
where $\Delta t$ is the time-step size, and 
\beq \lb {eq.RHS}
\textrm{RHS}' = - \widetilde{\po} \times \widetilde{\pu} + {\pf}_{\rm pre} - \na \cdot \pt, \quad
\textrm{RHS} = \textrm{RHS}' - \na \widetilde p.
\eeq
Then the next-step velocity $\widetilde \pu^{t+\Delta t}$ is obtained by 
\beq \lb {eq.u-t+1}
\frac{ \widetilde \pu^{t+\Delta t} - \widetilde \pu^* }{ \Delta t } = - \frac{3}{2} \na \widetilde {p}^t.
\eeq 
From Eq.~\er{eq.u-t+1} and the divergence-free condition, the pressure satisfies the Poisson equation,
\beq \lb {eq.p-Poisson}
\nabla^2 \widetilde {p}^t = \frac{2}{3\Delta t} \na \cdot \widetilde \pu^{*}.
\eeq
To obtain the pressure field, Eq.~\er{eq.p-Poisson} is written with $x$ and $y$ directions represented spectrally, while the vertical $z$ direction is kept in general operator form
\beq \lb {eq.p-Poisson-Fourier}
\left[ \pat_z^2 - \left( k_x^2 + k_y^2 \right) \right] \widehat p = \frac{2}{3\Delta t} \left[ i k_x \widehat {u^*} + i k _y \widehat {v^*} + \pat_z \widehat {w^*} \right].
\eeq 
Boundary conditions for Eq.~\er{eq.p-Poisson-Fourier} are available from consideration of the vertical momentum balance across the upper and lower boundaries, 
\beq\lb{eq.p-bc}
\pat_z \widehat {p} = \widehat {f}_z - \widehat { \pat_j \tau_{jz} }, \quad \mbox{at $z=0$ and $z= L_z$}.
\eeq

\section{Wall modeled immersed boundary method}
\label{sec.immersed boundary method}

The advantage of the wall modeled immersed boundary method is that the large-eddy simulation flow solver described above can be used without any adaptation. Also, it allows us to consider arbitrarily complex geometries within a uniform framework. In comparison to a method that uses terrain-following coordinates, a downside of the immersed boundary approach is that a relatively high resolution is required to capture the flow near smooth, spherical like, objects. As the grid cells do not necessarily coincide with the object boundaries, the sub-grid scale quantities need to be modeled carefully close to the immersed boundary \citep{cri06, yan15c}. Here, we apply the direct forcing approach \citep{moh97, fad00} to ensure that the velocity within the solid domain is always zero. To simplify the modeling of the sub-grid scale stresses, we use the Smagorinsky model \citep{sma63}. In Section \ref{sec.distance function} we introduce a signed distance function, which is used to implement the wall modeled immersed boundary method efficiently. In Section \ref{sec.ibm force} we discuss the treatment of the immersed boundary, and in Section \ref{sec.wall unsolved} the wall-stress modeling.

\subsection{Signed distance function} \lb{sec.distance function}
A signed distance function (or level-set function), $\phi$, is used to distinguish the fluid and solid regions,
\beq 
\phi(\px) = \left\{ 
\begin{split}
	& +d, & \quad & \px \in \textrm{fluid region}, \\
	& -d, & \quad & \px \in \textrm{solid region}, \\
	& 0, & \quad & \px \in \textrm{immersed boundary}, \\
\end{split}
\right.
\eeq
where $d$ is the minimum distance from the location $\px$ to the immersed boundary. Then, the normal unit vector to the boundary, $\pn$, can be computed as
\beq
\pn = \frac{1}{| \na \phi |} \na \phi.
\eeq
For a given terrain geometry, $\phi$ and $\pn$ are computed only once. 

\subsection{Immersed boundary approach} \lb{sec.ibm force}
The immersed boundary force is applied to the intermediate velocity instead of the physical velocity to ensure that the resulting velocity field is divergence-free \citep{fad00, bal04}. This means the pressure equation is still treated in the same way over the entire computational domain. This is based on the assumption that the direct forcing approach does not directly affect the pressure. Instead, the pressure field adjusts itself automatically through the velocity values imposed in the solid body and through the Poisson equation solved within the projection method \citep{fad00, tyl18}. 

To achieve this the intermediate velocity $\pu^{*}$ is obtained by solving Eq.~\er{eq.u-inter} without considering the immersed boundary force $\pf_{\rm IB}$. Then $\pu^{*}$ is updated by adding the force $\pf_{\rm IB}$,
\beq \lb{eq.u-inter-IB}
\frac{ \widetilde \pu^{*}_\textrm{IB} - \widetilde \pu^{*}}{ \Delta t } = \pf_{\rm IB},
\eeq
where 
\beq \lb {eq.f-IB}
\pf_{\rm IB} (\px) = \left\{ 
\begin{split}
	& \bf 0, & \textrm{if} \ \phi(\px) > 0, \\
	& \frac{3}{2} \na \widetilde {p}^{t-\Delta t} - \frac{\widetilde \pu^{*} }{ \Delta t }, & \textrm{if} \ \phi(\px) \le 0.
\end{split}
\right.
\eeq 
Subsequently, the velocity at the next time step $\widetilde \pu^{t+\Delta t}$ is obtained by 
Eqs.~\er{eq.u-t+1} and \er{eq.p-Poisson} with $\widetilde {\pu}^{*}$ replaced by $\widetilde {\pu}^{*}_\textrm{IB}$. Thus, the implementation of the immersed boundary force requires minimal adaptations to the actual solver. 

Note that $\pf_{\rm IB}$ does not have to be evaluated explicitly in the simulations, because $\widetilde {\pu}^{*}$ can be directly updated as follows
\beq \lb {eq.u-*-immersed boundary method}
\widetilde {\pu}^{*}_\textrm{IB} (\px) = \left\{ 
\begin{split}
	& \widetilde {\pu}^{*} (\px), & \textrm{if} \ \phi(\px) > 0, \\
	& \frac{3 \Delta t}{2} \na \widetilde {p}^{t-\Delta t}, & \textrm{if} \ \phi(\px) \le 0,
\end{split}
\right.
\eeq 
which can be verified easily by substituting Eq.~\er{eq.f-IB} into Eq.~\er{eq.u-inter-IB}.

We remark that the resultant velocity $\widetilde \pu^{t+\Delta t}$ obtained by applying the immersed boundary force $\pf_{\rm IB}$ to the intermediate velocity ${\pu}^{*}$ is always zero inside the body. It turns out that with this treatment the velocity inside the body is always zero and the boundary condition Eq.~\er{eq.bc-bottom-1} is still satisfied. The main reason is that the velocity modification produced by the projection step, i.e., Eq.~\er{eq.u-t+1}, is always negligibly small when compared to the velocity values themselves. This treatment has been widely used, see for example \citet{fad00, bal04, tyl18, che07, li16}. A more detailed discussion about this treatment can be found in the Appendix of \citet{fad00}. 

\subsection{Wall-stress modeling}\lb{sec.wall unsolved}

\begin{figure}[tb!] 
	\centering
	\includegraphics[width=0.5\textwidth]{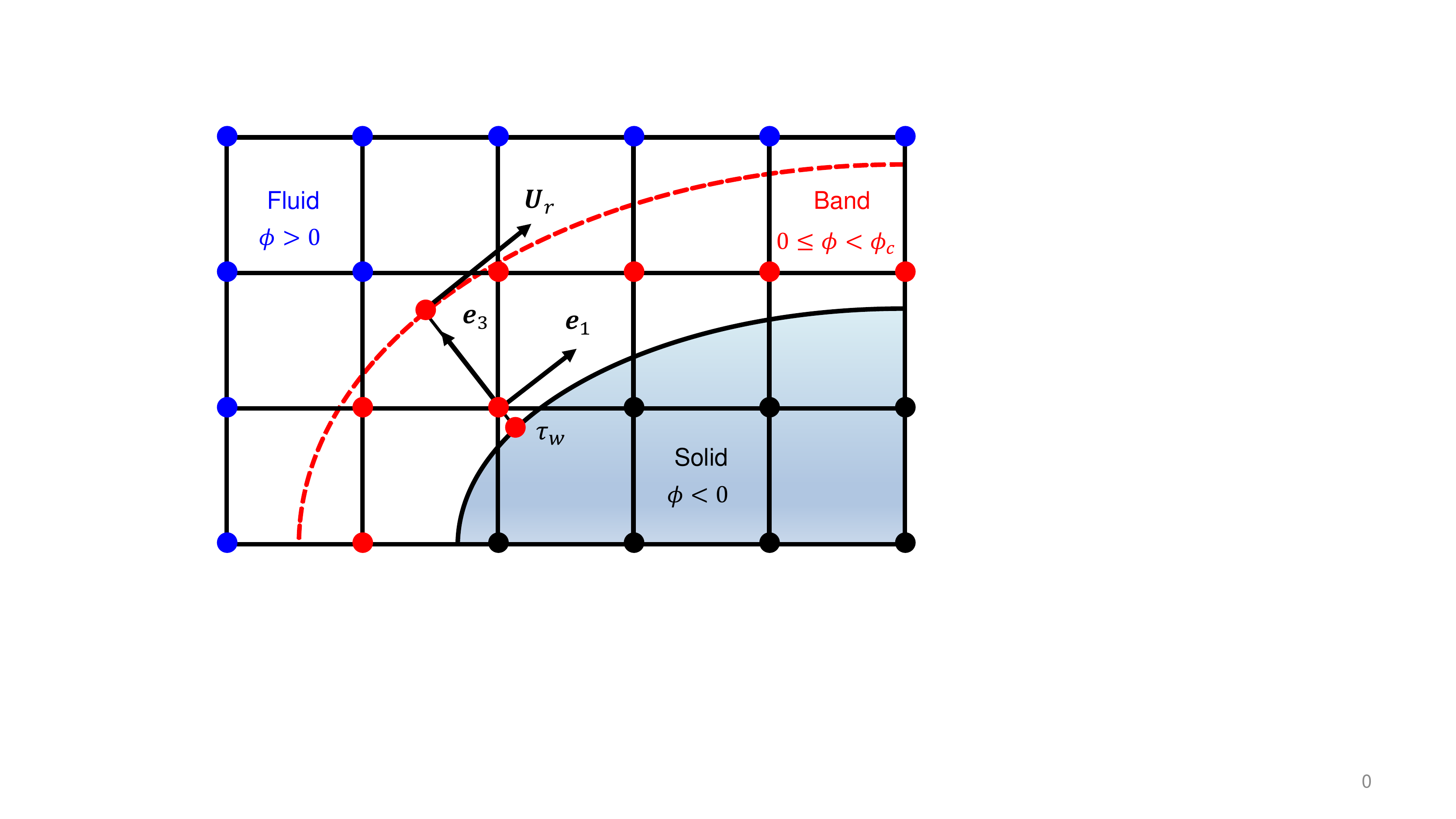}
	\caption{Sketch of the different regions used in the modeling of the sub-grid scale stresses near the wall, see Section~\ref{sec.wall unsolved} for details.}
	\label{fig.Chester}
\end{figure}

The resolution requirements for wall-resolved large-eddy simulation of high Reynolds number turbulent flows would lead to an unacceptably high computational cost. Therefore, following \citet{che07}, we use a wall-model, which relates the wall-stress with the tangential velocity in the vicinity of the immersed boundary surface \citep{pio02, che07}. We do this in a similar way as for the lower boundary, see Eq.~\er{eq.bc-bottom}. An efficient implementation is obtained by categorizing all points in the computational domain into three classes; see also Fig.~\ref{fig.Chester}:

\begin{itemize}
	\item $w$-grid
	\begin{itemize}
		\item[(1)] Fluid nodes: $\phi > \phi_b$. No further treatment. 
		\item[(2)] Band nodes: $-\phi_b \le \phi \le \phi_b$. The sub-grid scale stress is modeled. 
		\item[(3)] Solid nodes: $\phi < -\phi_b$. The sub-grid scale stress is set as zero. 
	\end{itemize}
	\item $uv$-grid
	\begin{itemize}
		\item[(1)] Fluid nodes: $\phi > 2\phi_b$. No further treatment. 
		\item[(2)] Band nodes: $0 \le \phi \le 2\phi_b$. The sub-grid scale stress is modeled. 
		\item[(3)] Solid nodes: $\phi < 0$. The sub-grid scale stress is set as zero. 
	\end{itemize}
\end{itemize}
Here, $ 2\phi_b \in [\Delta z, 1.5 \Delta z)$ is the size of the band region.
The choice of the band regions indicated above ensures that the wall-normal stresses $\tau_{xz}$ and $\tau_{yz}$ are set at the $w$-grid point closest to the immersed boundary where the vertical velocity $w$ should be zero. The other stress components $\tau_{xx}$, $\tau_{xy}$, $\tau_{yy}$ and $\tau_{zz}$, which correspond to the modeling of the horizontal velocity components $u$ and $v$, are determined at the first $uv$-grid point in the fluid domain. 
The detailed procedure to model the stresses is given below.

Let $\pn$ denote the surface normal passing through any band node $\px$ and $\pu$ the fluid velocity at a distance $\phi_c \ge 2\phi_b$ (see Fig.~\ref{fig.Chester}) above the solid surface along the line passing through $\px$ and parallel to $\pn$. Then $\pu$ and $\pn$ define a local coordinate system with origin at $\px$ and directions $\pe_1 = \pU_r$, $\pe_2 = \pn \times \pU_r$, and $\pe_3 = \pn$, with $\pU_r = \pu - \pu \cdot \pn \pn$. The velocity $\pu$ at this point is obtained by a 2nd-order accurate interpolation. Similar to the modeling used in Eq.~\er{eq.bc-bottom} the tangential velocity magnitude in this coordinate frame is assumed to follow an instantaneous logarithmic profile \citep{che07},
\beq\lb{eq.bc-bottom-immersed boundary method}
\tau_w = - \left[ \frac{ \kappa U_r }{ \ln (\phi_c/z_{0, \rm IB})} \right]^2,
\eeq
where $z_{0, \rm IB}$ is the roughness height of the wall surface, and $U_r$ is magnitude of the tangent velocity $\pU_r$. Even though $\px$ may not be exactly on the wall itself, the shear stress $\tau_w$, for simplicity, is taken to represent the stress components $\tau'_{13}(\px) = \tau'_{31}(\px)$ in the local coordinate frame. The reason is that this allows us to preserve the vertical stress-balance, see Eq.~\er{eq.lin-law-txz-uw} below, in the turbulent boundary layer. Note that in these simulations the viscous sublayer is not resolved. Instead, the velocities and velocity gradients near the wall are modeled. As a result it is not possible to perform consistent and accurate interpolations of all velocity gradients, which would be required to set all stress components. Therefore, to avoid reconstructing the velocity gradient near the wall \citep[e.g.][]{fan16}, we set the non-dominant stress components to zero. In this work, we show that the resulting scheme ensures that the vertical stress-balance in the boundary layer is preserved, while good agreement between the simulation results and wind tunnel measurements is obtained.

Finally, the stress tensor in the local coordinate system is transformed (rotated) to the global coordinate system to set the appropriate components of the stress tensor $\pt(\px)$. The tensor rotation is carried out using $\tau_{ij} = a_{im} a_{jn} \tau'_{mn}$, where $a_{ij}$ are the directional cosines between the global bases $(\pe_x,$ $\pe_y,$ $\pe_z)$ and the rotated bases $(\pe_1, \pe_2, \pe_3)$.

To summarize, the present method is developed from the smearing method proposed by \citet{che07} but has two distinct features. First, we treat the $w$-grid and $uv$-grid points separately while \citet{che07} treat them in the same way. Second, we set the stresses inside the body to zero directly while \citet{che07} uses a linear extrapolation to obtain the stresses in the solid band region and a successive over-relaxation (SOR) iteration method to smooth the stresses further inside the body. The present particular way to reconstruct the stresses near the immersed boundary allows us to reconstruct the velocity profile accurately and preserve the vertical stress balance in all situations as is explained for the flat terrain test cases in Section~\ref{sec.flat} of the manuscript.

A significant benefit of the proposed method is that it improves the computational efficiency compared to the \citet{che07} method, especially for highly parallel simulations. The reason is that it prevents the extrapolation (step 2) and SOR iteration steps (step 3) of the \citet{che07} method. Removing these steps reduces the number of computations. However, more importantly, removing the extrapolation step reduces the required communication and significantly improves the load balancing, which is very important in parallel computations. The observation that the computational overhead of the proposed method is less than $2\%$ underlines its suitability for parallel high-performance simulations. From a practical point of view, the proposed method is also attractive as it is easier to implement than the \citet{che07} method.

\section{Validation}
\lb{sec.applications} 

We validate the wall modeled immersed boundary method described above using various test cases. First, we consider a neutral atmospheric boundary layer over flat terrain. In Section~\ref{sec.flat} we evaluate the accuracy of the simplifications in the wall-stress modeling by elevating the lower wall in the computational domain and compare the results with the original reference case. Subsequently, we compare the simulation results with experimental measurements for flow over wall-mounted cubes (Section~\ref{sec.wall}), a two-dimensional ridge (Section~\ref{sec.hill-2d}), and a three-dimensional hill (Section~\ref{sec.hill-3d}).

\subsection{Flat terrain simulations}
\lb{sec.flat} 

\begin{figure}[tb!] 
	\centering
	\includegraphics[width=0.6\textwidth]{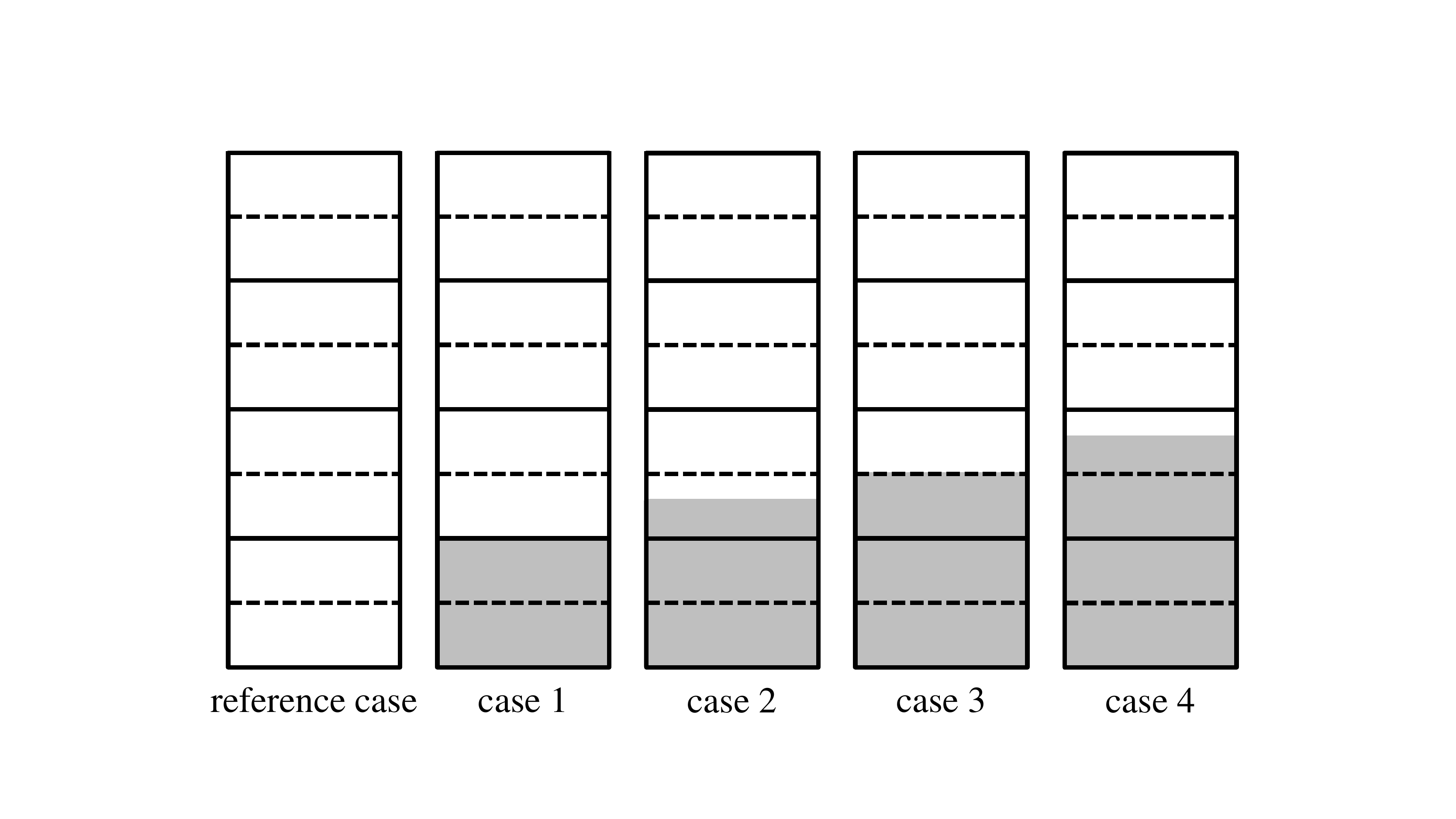}
	\caption{Sketch of the wall location, i.e.\ the shaded region, with respect to the grid locations (Solid line:\ $w$-grid; Dashed line:\ $uv$-grid) for the cases discussed in Section \ref{sec.flat}. The vertical location of the wall surface is $z_w / \Delta z = 0$ (reference case), $z_w / \Delta z = 1$ (case 1), $z_w / \Delta z = 1.25$ (case 2), $z_w / \Delta z = 1.5$ (case 3), $z_w / \Delta z = 1.75$ (case 4).}
	\label{fig.grid}
\end{figure}

To validate the wall modeled immersed boundary method we compare the results for flow over a flat plate. Four different cases are considered, in which the lower wall is located at $z_w / \Delta z = 1, 1.25, 1.5, 1.75$, respectively (cases $1$ to $4$ in Fig.~\ref{fig.grid}) when it is modeled with the wall modeled immersed boundary method. The results are compared with the traditional situation in which the wall is perfectly aligned with grid and located at $z_w = 0$ (the reference case in Fig.~\ref{fig.grid}). This allows us to evaluate the accuracy of the model for cases in which the wall is not aligned with the grid. The size of the simulation domain is $L_x \x L_y \x L_z = (8 \x 4 \x 1) L_z$ and the roughness height is $z_0/L_z = 5.6\x 10^{-5}$. The mean pressure gradient $\pf_{\rm pre}$ is set to $1/(L_z-z_w)$ such that the horizontally averaged wall-stress $\langle \tau_w \rangle = -1$. For each case two different grid resolutions are considered, i.e., $N_x \x N_y \x N_z = 64\x 32 \x 33$ and $256 \x 128 \x 129$.

\begin{figure}[tb!] 
	\centering
	\begin{overpic}[width=0.45\textwidth]{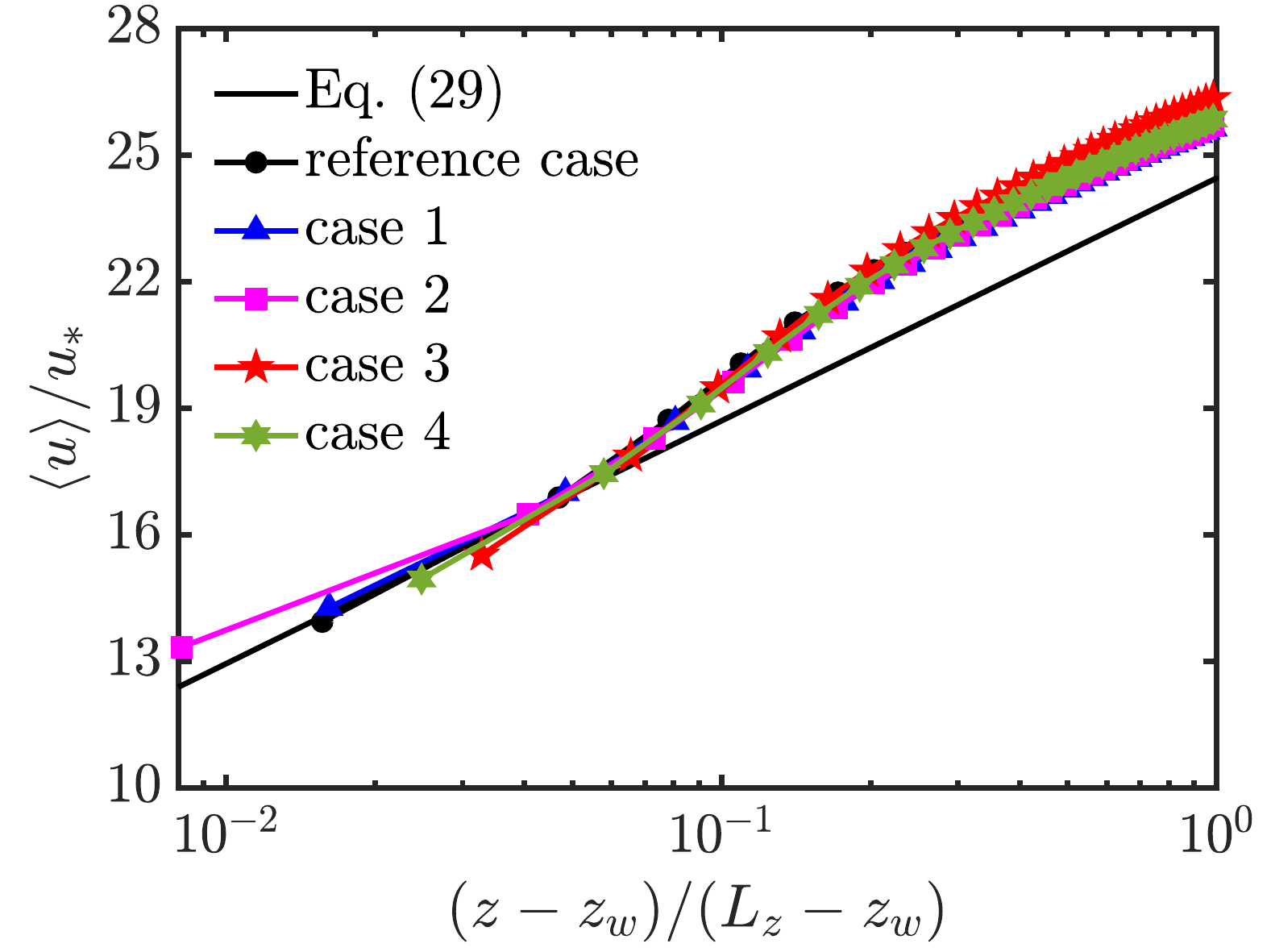} 
		\put(0,69){\footnotesize $(a)$}
	\end{overpic}
	\begin{overpic}[width=0.45\textwidth]{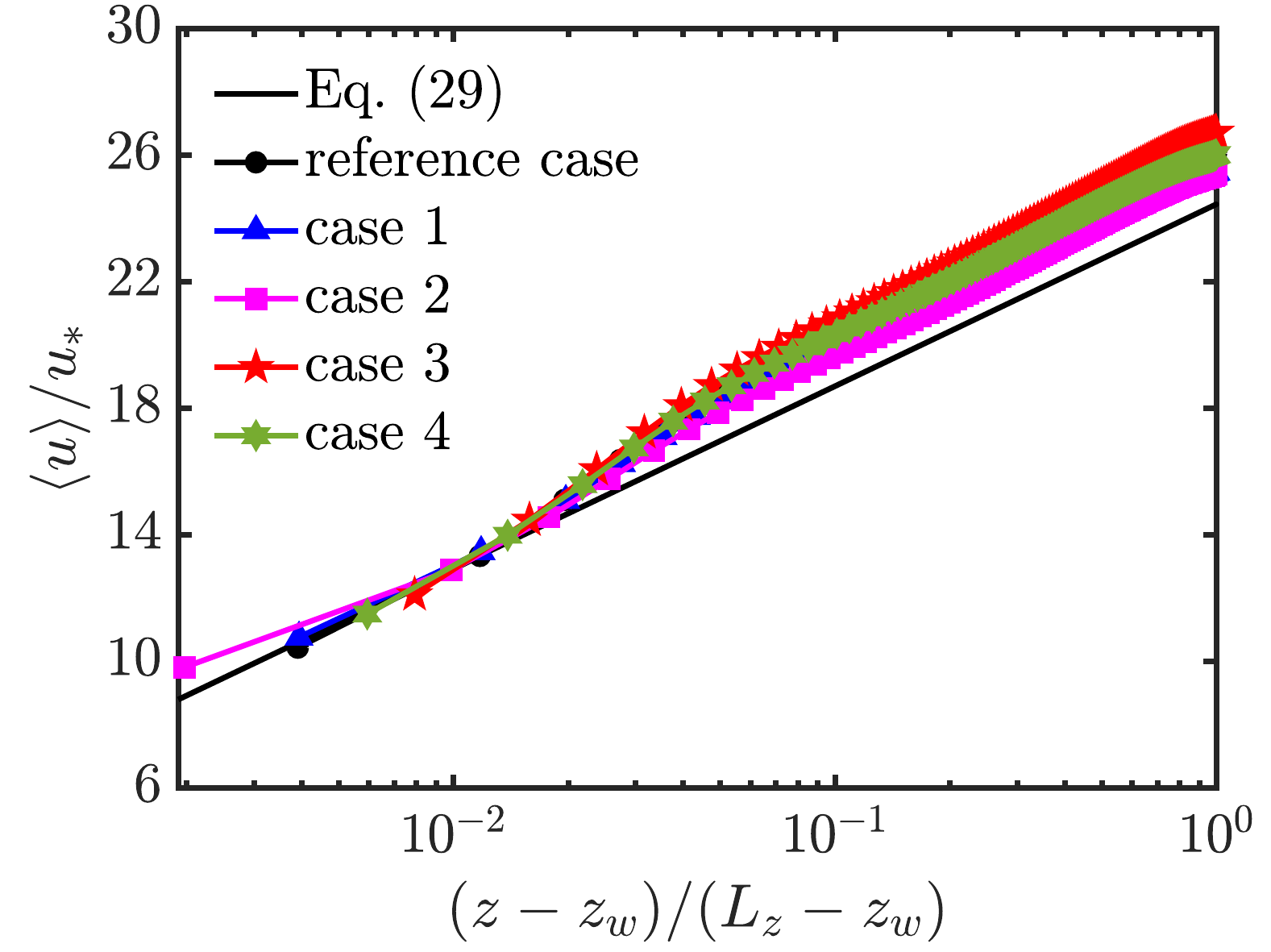} 
		\put(0,69){\footnotesize $(b)$}
	\end{overpic}
	\caption{Horizontally and time averaged streamwise velocity profiles for simulations over flat terrain performed on a (a) $64\x 32 \x 33$ and a (b) $256\x 128 \x 129$ grid. The different cases are sketched in Fig.~\ref{fig.grid}. }
	\label{fig.flat-u}
\end{figure}

Figure~\ref{fig.flat-u} compares the time-averaged streamwise velocity profiles from the different simulations with the logarithmic law for the streamwise velocity in a turbulent boundary layer, i.e.
\beq\lb{eq.log-law-u}
\fr{ \langle u \rangle}{u_*} = \fr{1}{\ka} \ln \left( \fr{z-z_w}{z_0} \right),
\eeq
where $\langle \cdot \rangle$ denotes time average, and $u_* = \sqrt{-\tau_w}$ is the friction velocity. The figure shows that for case 1 the results using the wall modeled immersed boundary method perfectly collapses with the reference large-eddy simulation results in which the lower wall is located at the $w$-grid. Some small, but unavoidable, differences between the reference solution and the other test cases are observed. These differences originate from the fact that the surface does not coincide with the grid. Hence the boundary conditions are not known at the grid locations but have to be imposed at some location in the fluid or inside the wall, which leads to unavoidable approximations.

\begin{figure}[tb!] 
	\centering
	\begin{overpic}[width=0.45\textwidth]{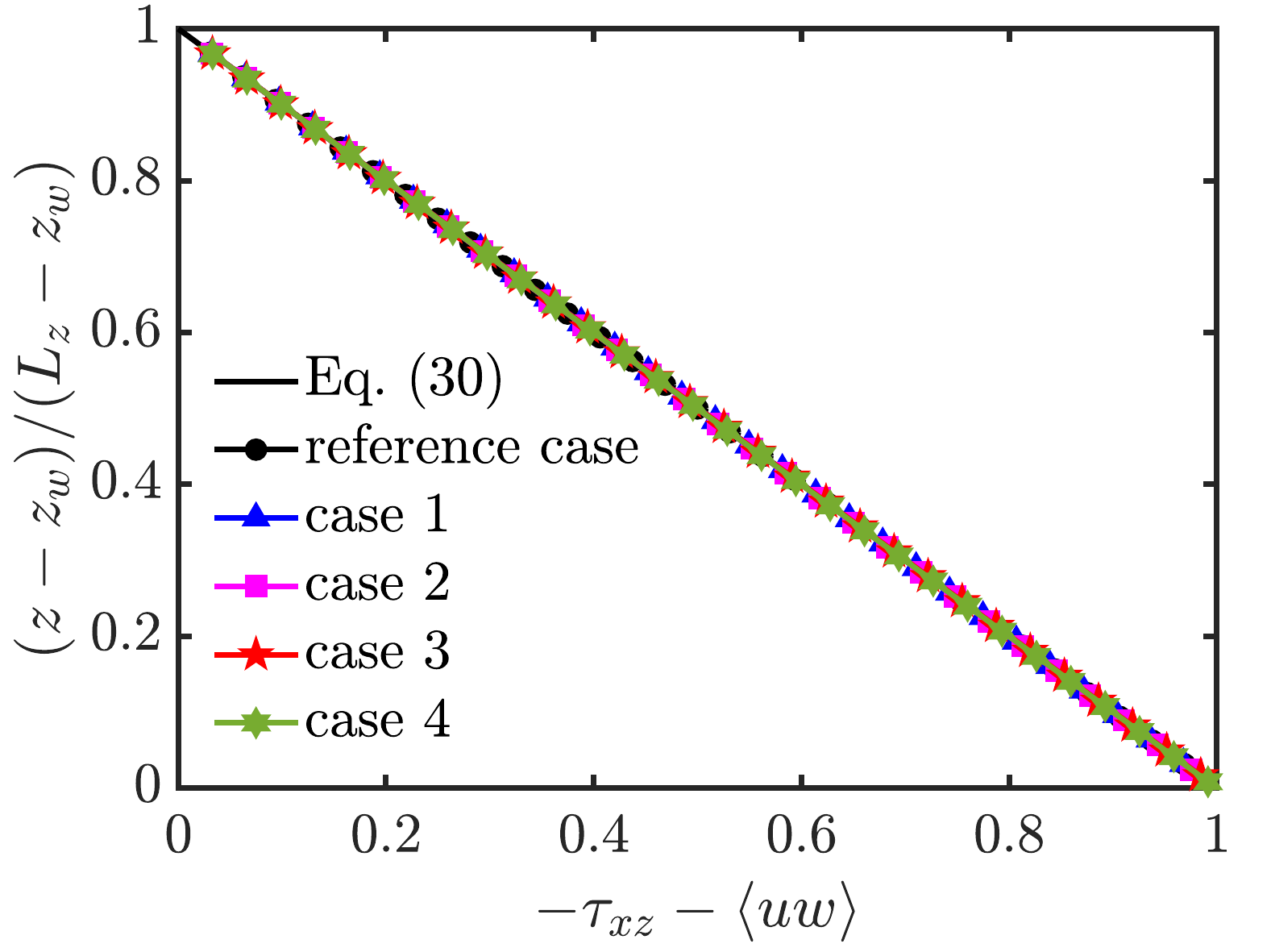} 
		\put(0,69){\footnotesize $(a)$}
	\end{overpic} 
	\begin{overpic}[width=0.45\textwidth]{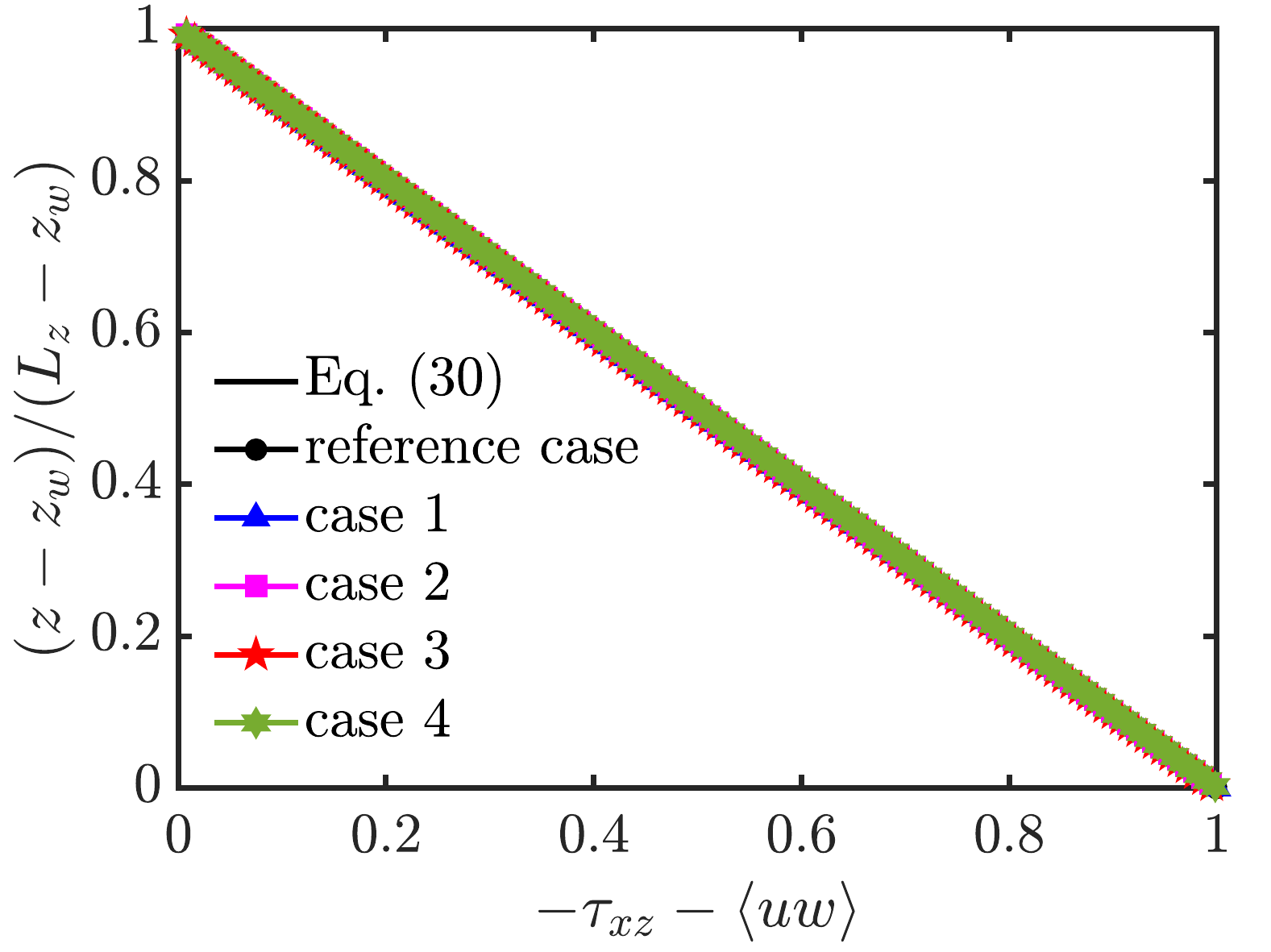} 
		\put(0,69){\footnotesize $(b)$}
	\end{overpic}
	\caption{Horizontally and time averaged stress profiles for simulations over flat terrain performed on a (a) $64\x 32 \x 33$ and a (b) $256\x 128 \x 129$ grid. The different cases are sketched in Fig.~\ref{fig.grid}. }
	\label{fig.flat-txz-uw}
\end{figure}

In turbulent boundary layer simulations, it is important to ensure that the total stress $-\tau_{xz} - \langle u w \rangle$ profile follows the theoretical result
\beq\lb{eq.lin-law-txz-uw}
-\tau_{xz} - \langle u w \rangle = 1 - \fr{z-z_w}{L_z-z_w}.
\eeq
Figure~\ref{fig.flat-txz-uw} shows that this important condition is perfectly satisfied for all cases. While some of the features discussed below can be improved, this is not possible without violating this requirement.

\begin{figure}[tb!] 
	\centering
	\begin{overpic}[width=0.45\textwidth]{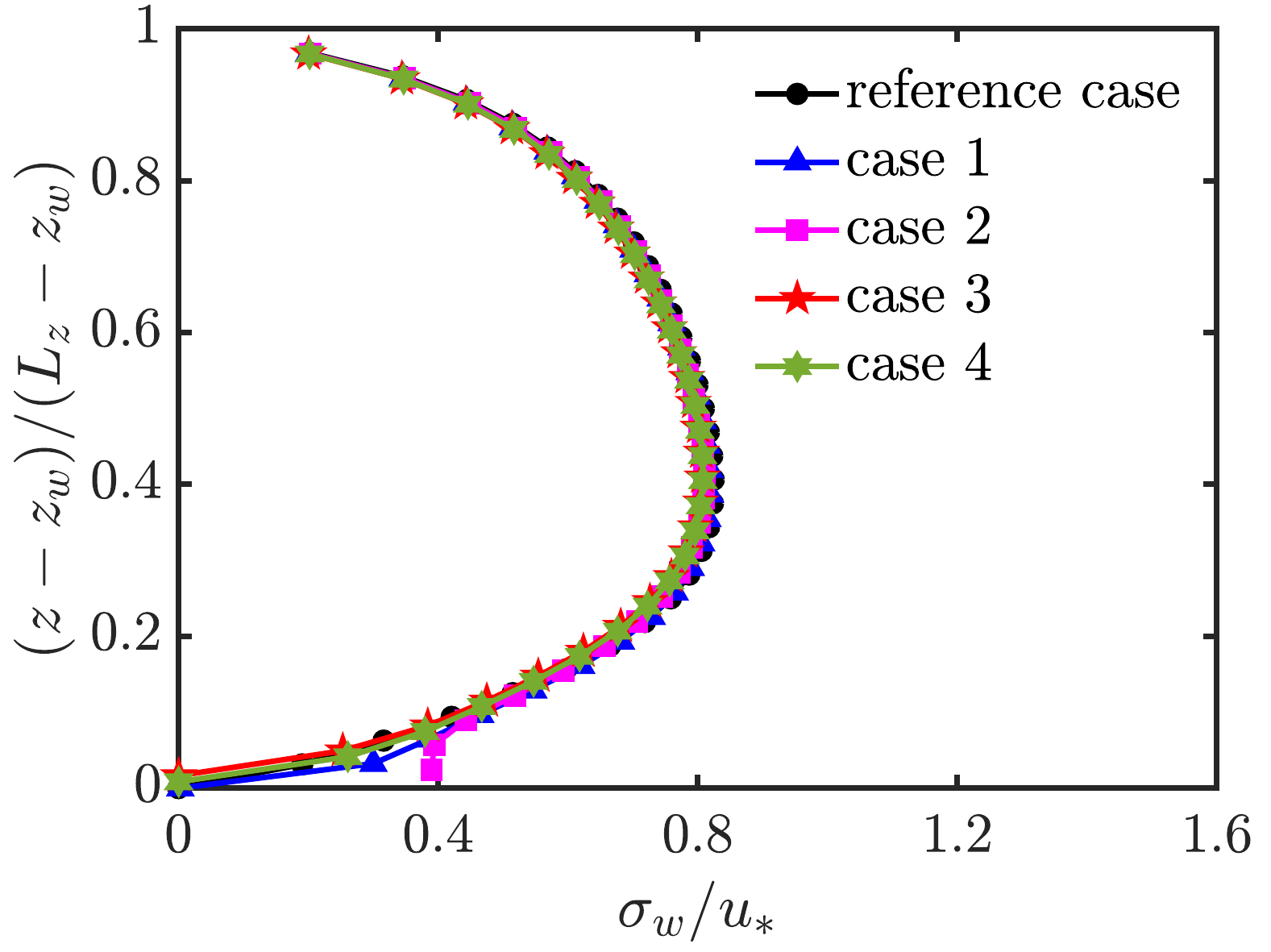} 
		\put(0,69){\footnotesize $(a)$}
	\end{overpic} 
	\begin{overpic}[width=0.45\textwidth]{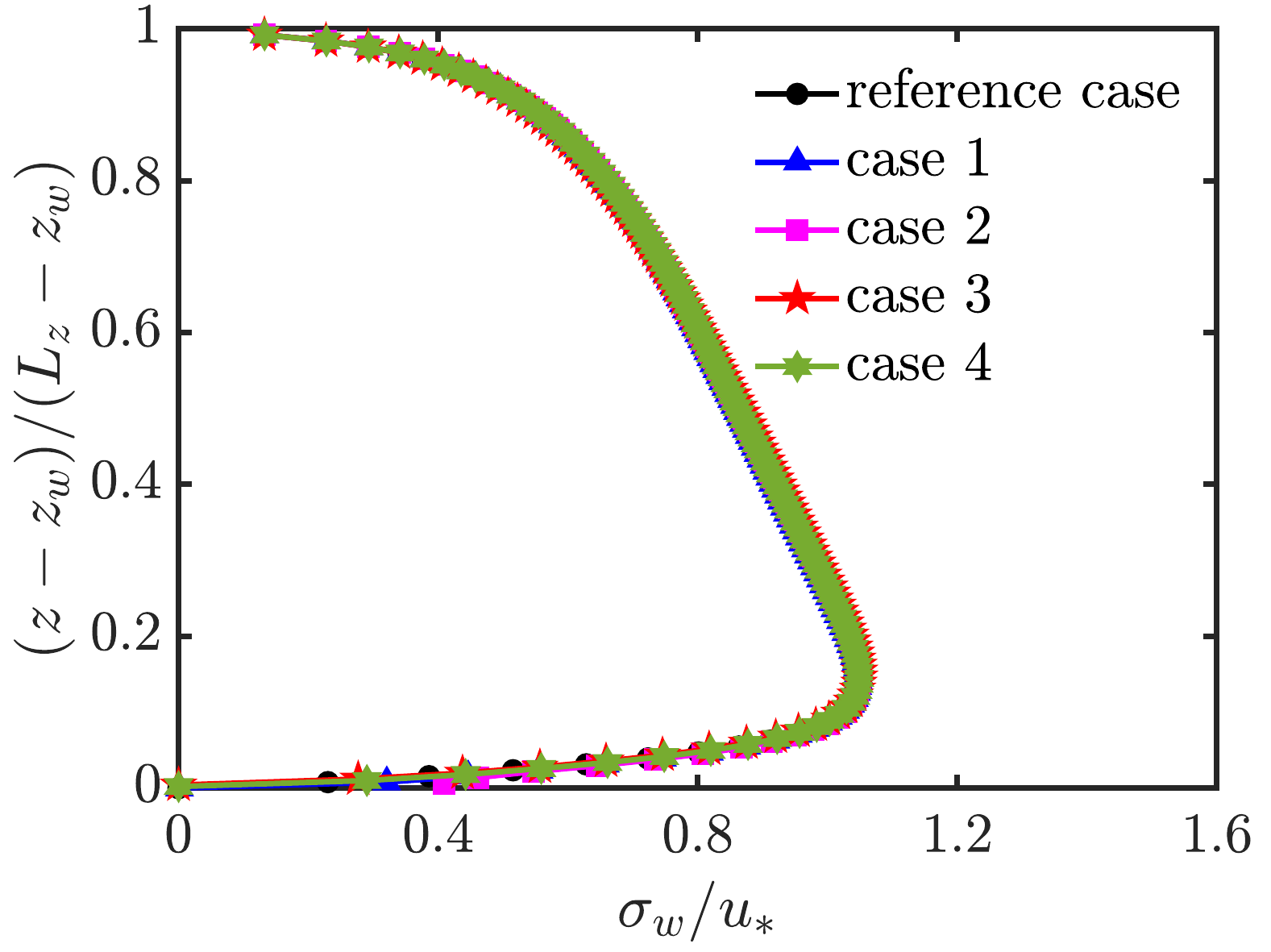} 
		\put(0,69){\footnotesize $(b)$}
	\end{overpic}
	\caption{Horizontally and time averaged vertical velocity variance profiles for simulations over flat terrain performed on a (a) $64\x 32 \x 33$ and a (b) $256\x 128 \x 129$ grid. The different cases are sketched in Fig.~\ref{fig.grid}. }
	\label{fig.flat-wrms}
\end{figure}

Figure~\ref{fig.flat-wrms} shows that the vertical velocity $w$ and its variance $\sigma_w= \sqrt{w^2}$ are zero at the first grid point for cases 3 and 4 when the first point in the fluid domain is a $w$-node instead of a $uv$-node. This is not physical because there must be some vertical velocity fluctuations in the fluid domain. However, it is a numerical consequence of imposing the divergence-free condition using the same treatment as if the solid body was not there. To be specific, for cases 3 and 4 the velocities and horizontal derivatives at the $uv$-node at or just below the surface is zero. The vertical derivative $\pat_z w$ is calculated according to Eq.~\er{eq.dwdz}, where both solid and fluid points are involved. Since $w$ and $\pat_z w$ are zero the vertical velocity at the first $w$-node inside the fluid domain becomes zero by imposing the divergence-free condition.

When the first grid node above the wall surface is a $uv$-node and the $w$-grid does not coincide with the wall (e.g., case 2), the streamwise velocity $u$ at the first fluid node is slightly above the predicted value (see Fig.~\ref{fig.flat-u}). The reason is that the wall sub-grid scale stress is stored on the closest $w$-node, but the calculation of the divergence of the sub-grid scale stresses is treated as if there is no solid body. The streamwise velocity prediction for the first point can be improved by performing a linear interpolation between the stresses at the wall and the stresses in the fluid region. However, using this procedure violates the total stress-balance, see Eq.~\er{eq.lin-law-txz-uw}, and is therefore not used.

\subsection{Complex terrain simulations: flow over wall mounted cubes} 
\lb{sec.wall}

To validate the method for a more complicated case, we consider the turbulent flow over wall-mounted cubes. The simulation results are compared with the data from wind tunnel measurements conducted by \citet{mei99}. The experiments were performed in a wind tunnel with a rectangular test section with a height of 0.6~m and a width of 0.051~m. A matrix of cubes with size $h=0.015$~m, spaced equidistantly at 0.045~m (face-to-face) in both the streamwise and spanwise direction, was mounted on one of the channel walls. The matrix consisted of a total of $25 \times 10$ cubes in the streamwise and spanwise directions, respectively. The measurements were performed around the 18th row counted from the inlet, where the flow is fully developed and nearly independent of the inflow conditions. The bulk velocity $u_B = 3.86$~m/s and the Reynolds number based on the bulk velocity and the cube height is $Re_h = 3854$. 

In the simulations we use a periodic domain with $2 \times 2$ cubes in the streamwise and spanwise direction, respectively, to model the experiments. The domain size is $L_x \x L_y \x L_z = 8h \times 8h \times 3.5h$, where $h$ is the cube height. Three different grid resolutions are considered, i.e.\ $64 \x 64 \x 29$, $128 \x 128 \x 57$, and $256 \x 256 \x 113$, such that $\Delta x=\Delta y = \Delta z$ and the top of the cubes aligns with the $w$-grid. The roughness height is set as $z_{0, \rm IB}/h = z_0/h = 5.7 \x 10^{-4}$. 

\begin{figure}[tb!] 
	\centering
	\begin{overpic}[width=0.45\textwidth]{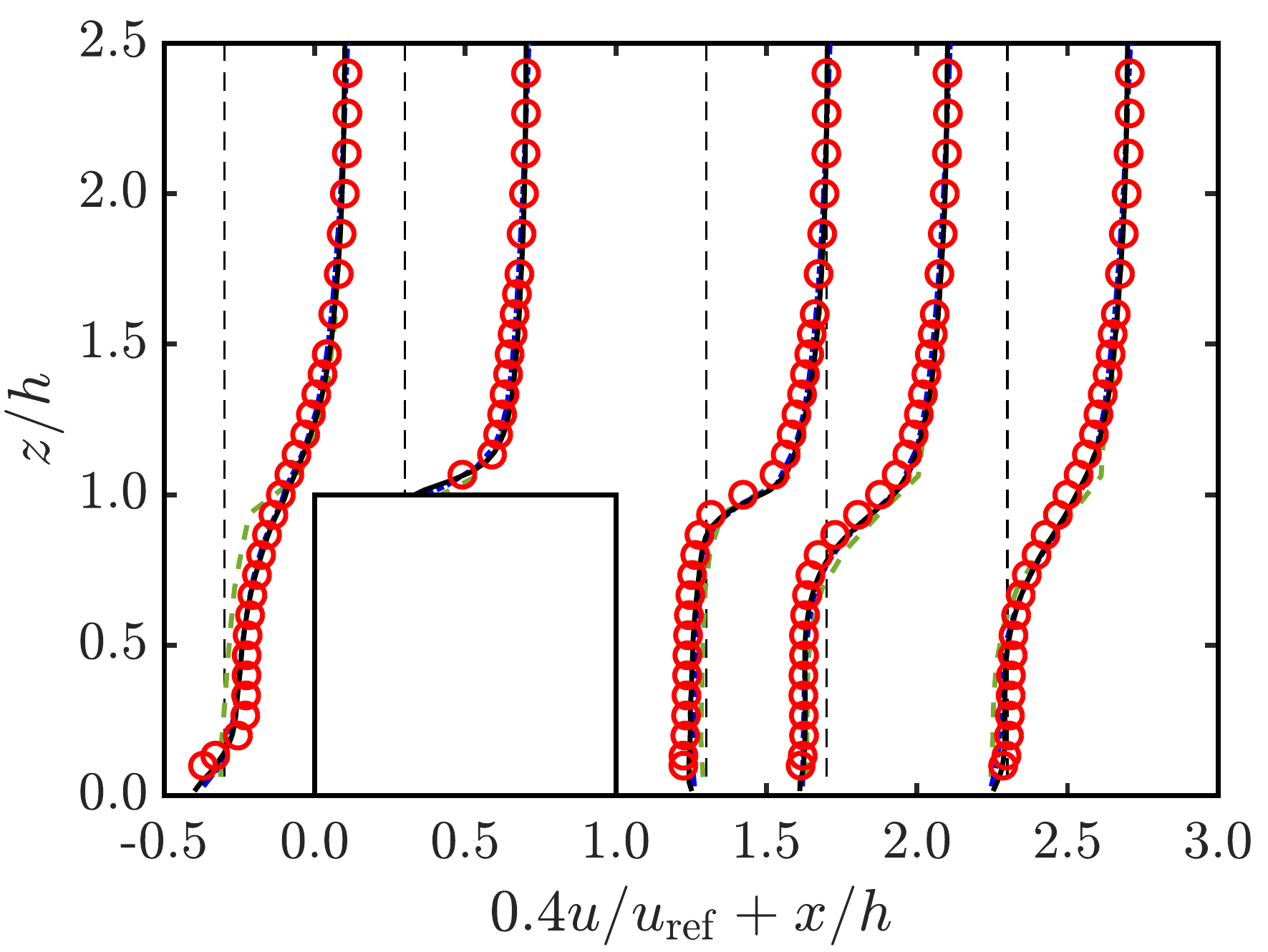} 
		\put(0,67){\footnotesize $(a)$}
	\end{overpic} \\
	\vspace{2mm}
	\begin{overpic}[width=0.45\textwidth]{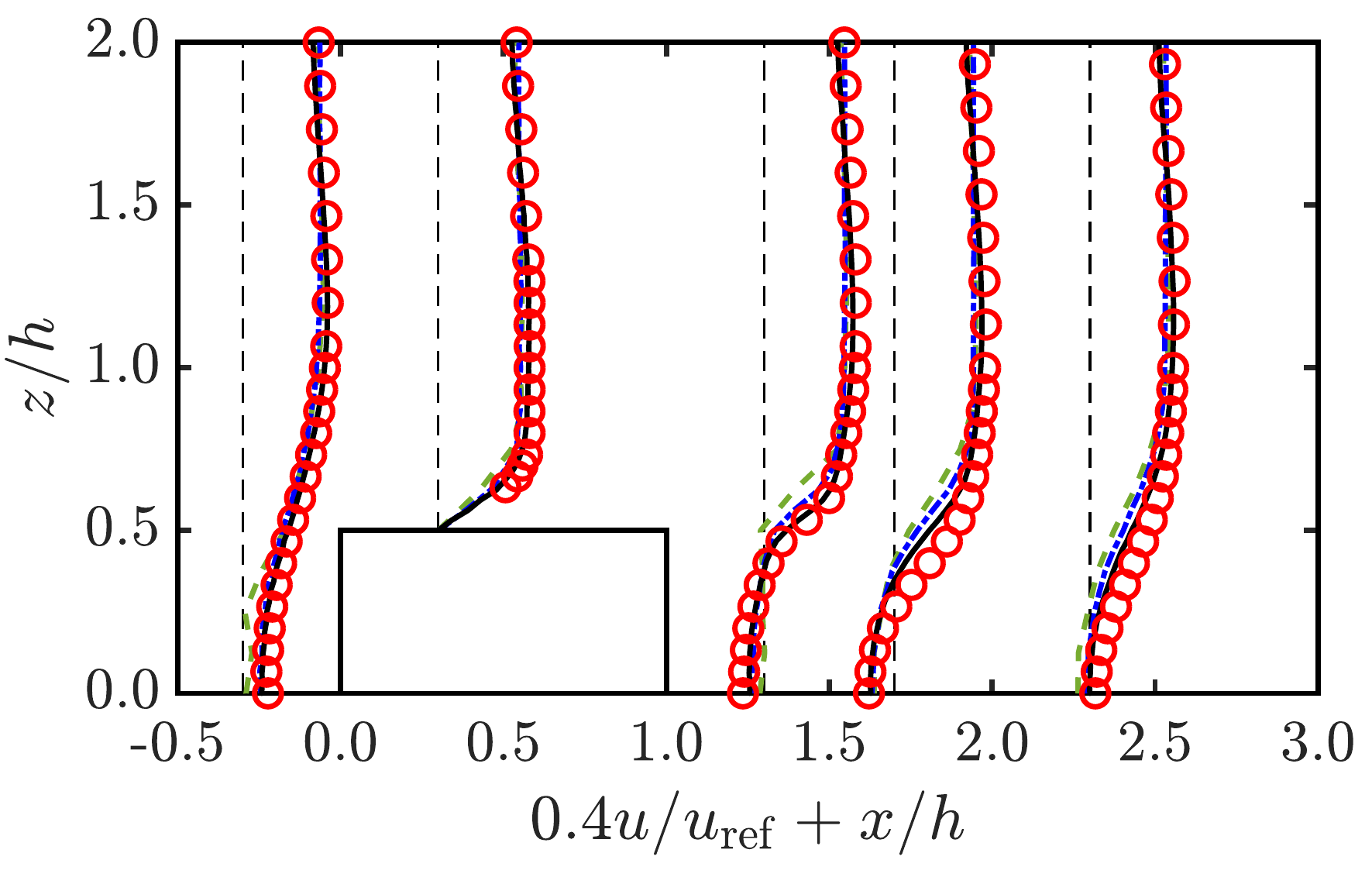} 
		\put(0,56){\footnotesize $(b)$}
	\end{overpic}
	\begin{overpic}[width=0.45\textwidth]{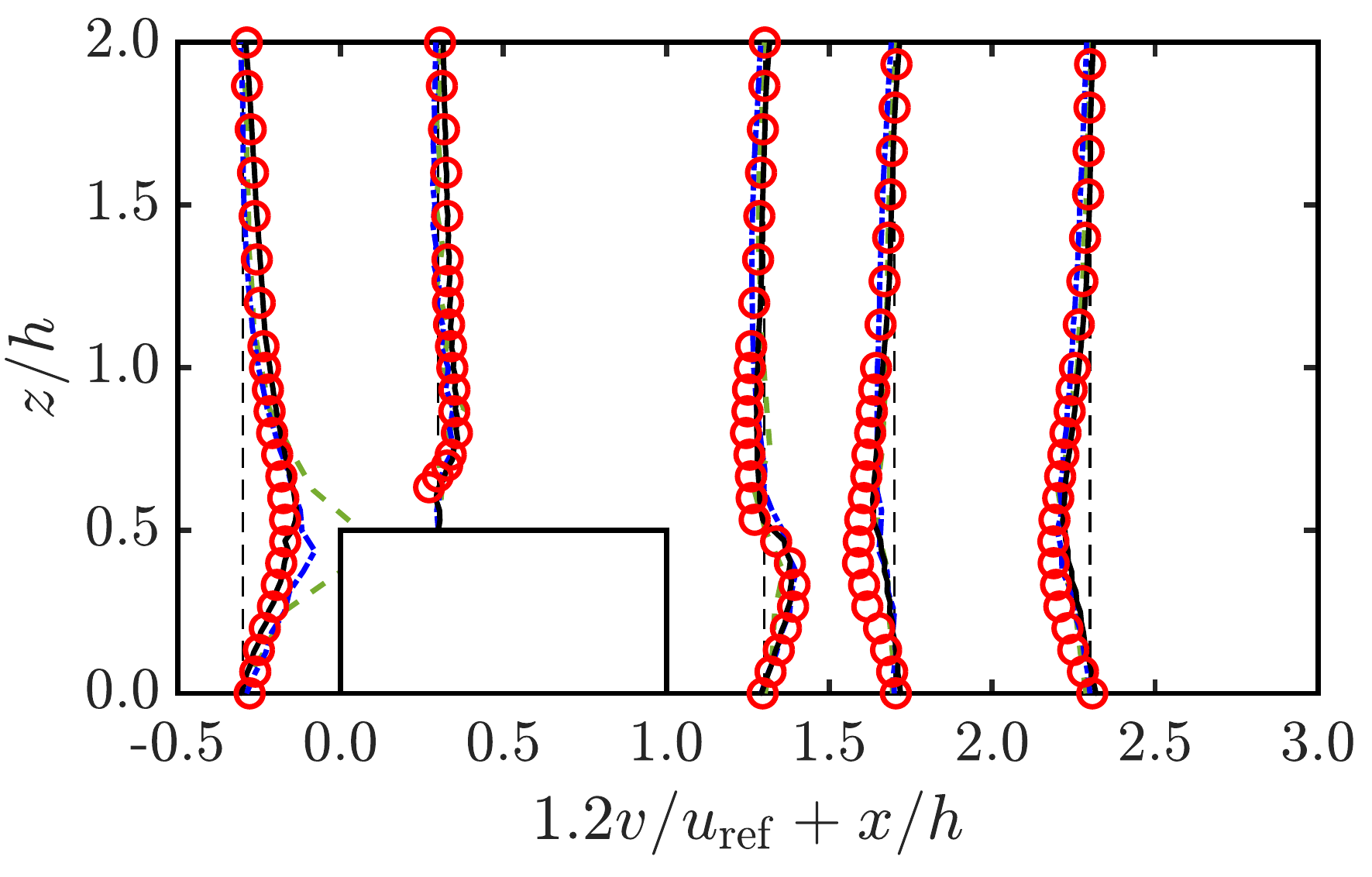} 
		\put(0,56){\footnotesize $(c)$}
	\end{overpic}
	\caption{Time averaged (a) vertical and (b) spanwise profile of the streamwise velocity and (c) the spanwise profile of the spanwise velocity for the turbulent flow over wall mounted cubes. Open circles: experimental data by \citet{mei99}; green dashed lines: $64\x 64 \x 29$; blue dashed-dotted lines: $128\x 128 \x 57$; black solid lines: $256\x 256 \x 113$. The vertical dashed lines indicate the measurement positions.}
	\label{fig.cube-u}
\end{figure}

\begin{figure}[tb!] 
	\centering
	\begin{overpic}[width=0.45\textwidth]{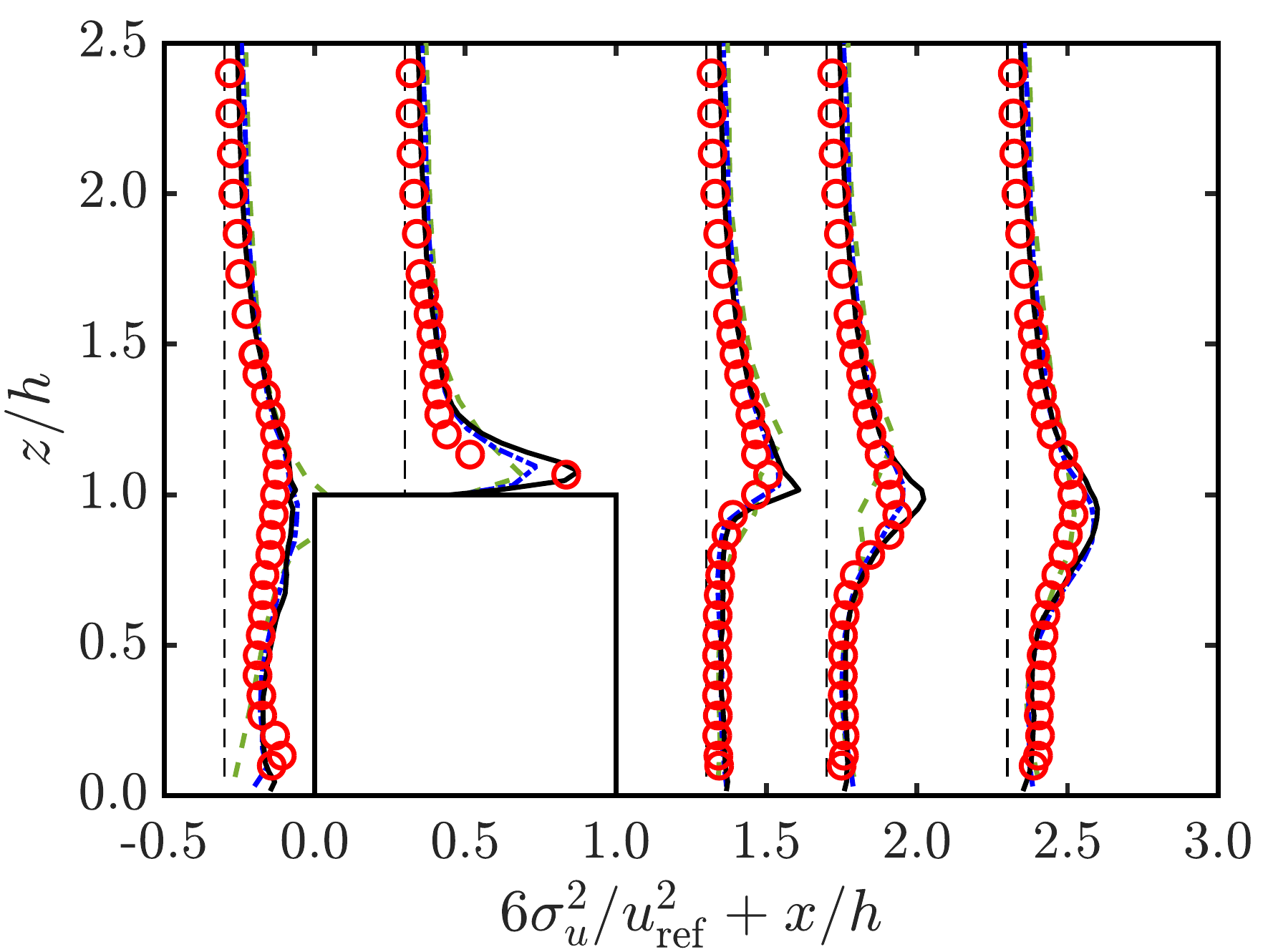} 
		\put(0,67){\footnotesize $(a)$}
	\end{overpic}
	\begin{overpic}[width=0.45\textwidth]{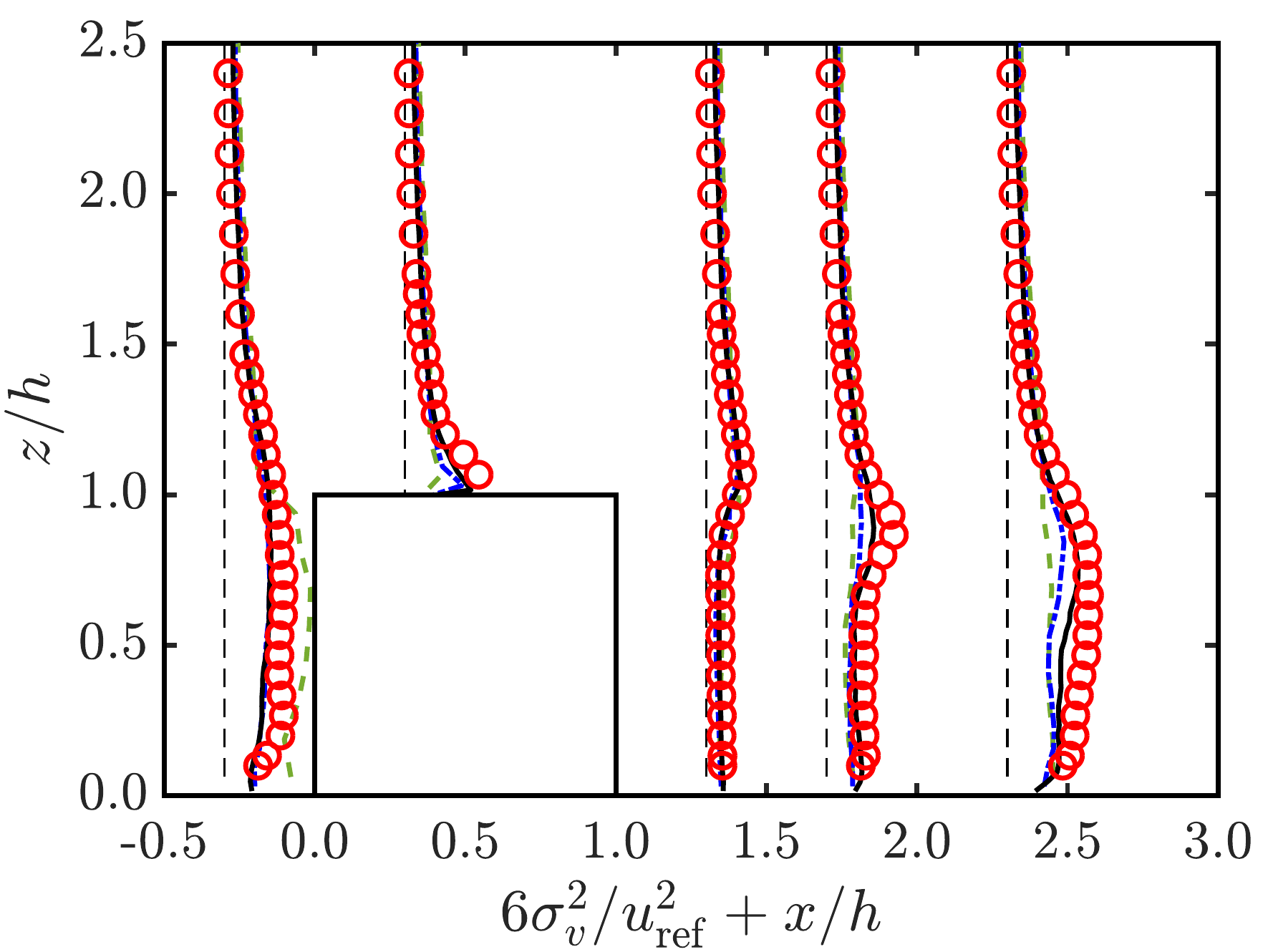} 
		\put(0,67){\footnotesize $(b)$}
	\end{overpic} \\
	\begin{overpic}[width=0.45\textwidth]{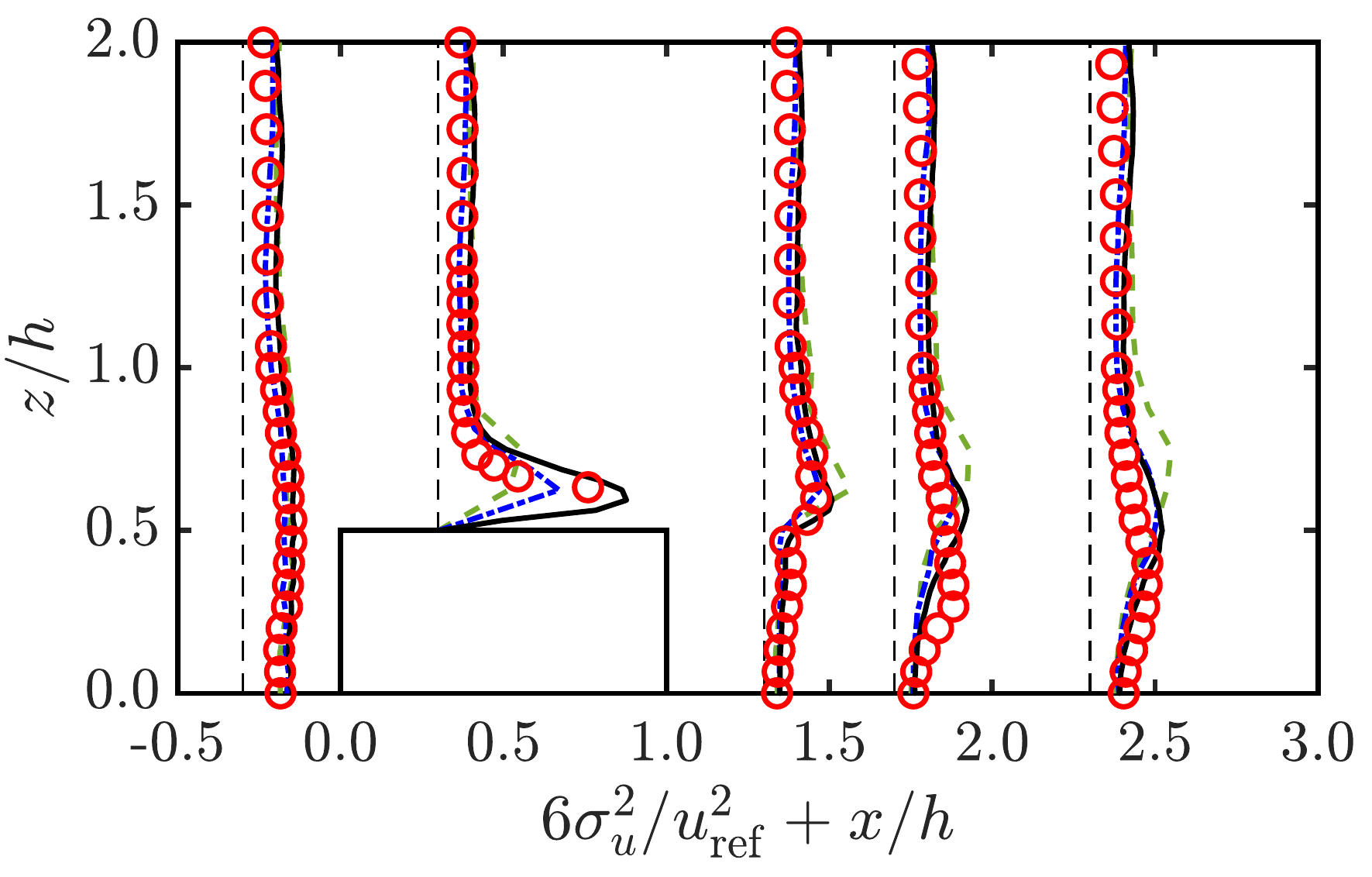} 
		\put(0,56){\footnotesize $(c)$}
	\end{overpic}
	\begin{overpic}[width=0.45\textwidth]{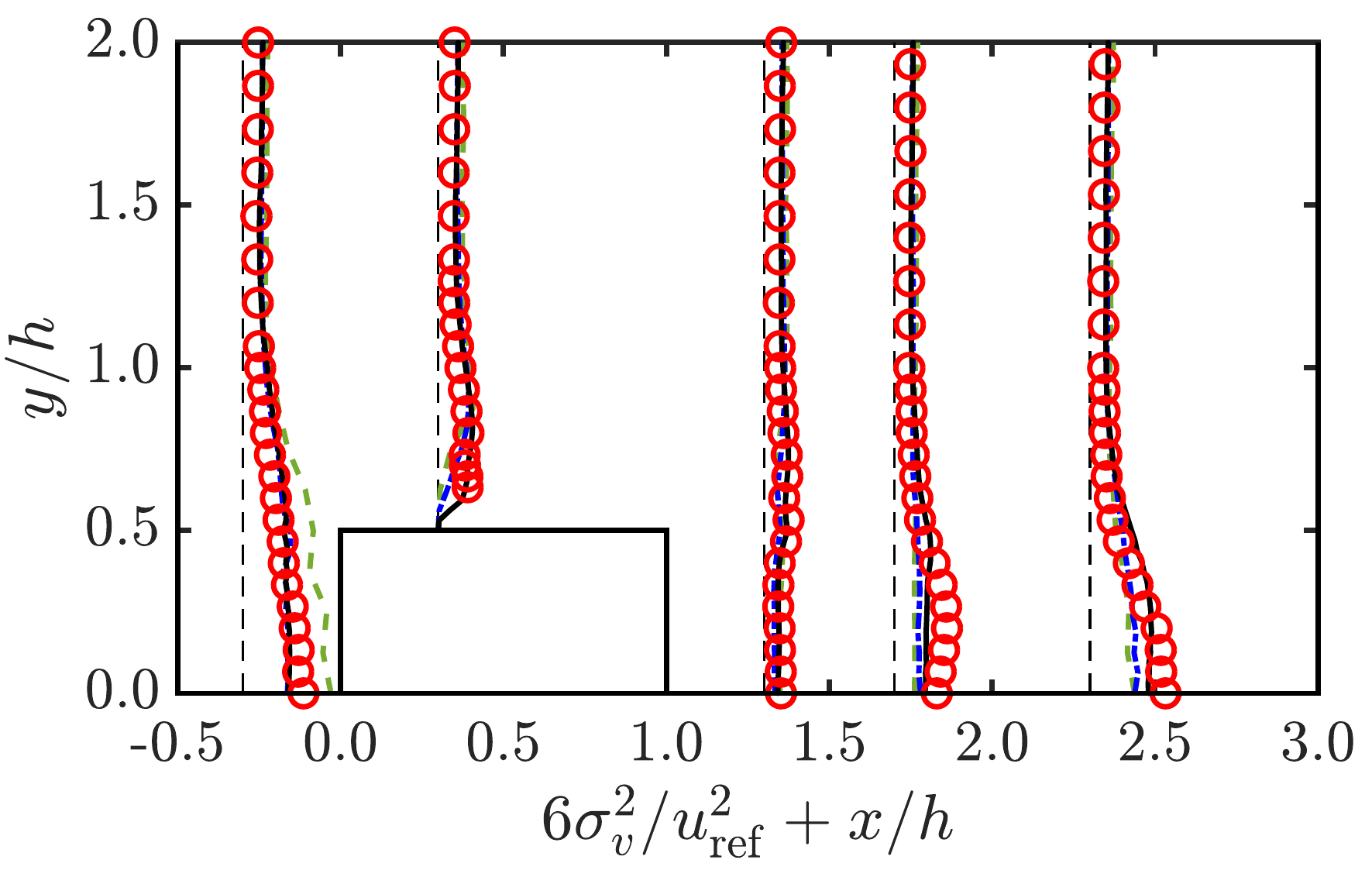} 
		\put(0,56){\footnotesize $(d)$}
	\end{overpic}
	\caption{Time averaged vertical profile of the (a) streamwise and (b) spanwise velocity variance and the corresponding spanwise profiles for the (c) streamwise and (d) spanwise velocity variance for the turbulent flow over wall mounted cubes. Open circles: experimental data by \citet{mei99}; green dashed lines: $64\x 64 \x 29$; blue dashed-dotted lines: $128\x 128 \x 57$; black solid lines: $256\x 256 \x 113$. The vertical dashed lines indicate the measurement positions.}
	\label{fig.cube-uu}
\end{figure}

Figure \ref{fig.cube-u} shows that the time-averaged velocity profiles obtained from the simulations agree excellently with the experimental measurements. In each panel data for five different measurement locations $(x/h = -0.3, 0.3, 1.3, 1.7, 2.3)$ is displayed. The local coordinate frame is defined such that $z/h=0$ indicates the streamwise position of the leading edge of the cube. All velocities are normalized using a reference velocity $u_{\rm ref}$ located at $(x, y, z)=(1.3h, 0, 2.25h)$ in the local coordinate frame of the cube. Figure \ref{fig.cube-uu} shows that also the corresponding Reynolds stresses agree very well with the experimental results, especially for the higher resolutions. We remark that a higher grid resolution is required to capture the Reynolds stress profiles than to capture the mean velocity profiles, see Fig.~\ref{fig.cube-u} and \cite{tse06, gra12, yan15}. It is a well-known feature of large-eddy simulation that the prediction of higher-order statistics requires a higher grid resolution, see, e.g., \citet{ste14d}.

\subsection{Complex terrain simulations: flow over two-dimensional ridge}
\lb{sec.hill-2d}

To evaluate the performance of the method for a more complex case we compare the simulations with the wind tunnel measurements by \citet{cao06} for flow over a two-dimensional cosine-squared ridge. The height profile $z_w$ of the two-dimensional ridge used in this experiment is described by 
\beq\lb{eq.hill-2d}
z_w(x) = h \cos^2 \left( \fr{\pi x}{2 b} \right), \quad -b \le x \le b,
\eeq
where $h=0.04$~m and $b=0.1$~m are the height and half-width of the ridge, respectively. The wind profiles were measured at nine different locations, i.e., $x/h=-2.5, -1.25$, 0, 1.25, 2.5, 3.75, 5, 6.25, 7.5. The flow measured without the ridge corresponded to a neutral atmospheric boundary layer with thickness $\delta = 0.25$~m. The friction velocity and the roughness height are $u_*/U_\infty = 0.03$ and $z_0/\delta = 1.6 \times 10^{-5}$, respectively. Here, $U_\infty$ is the free stream velocity outside the boundary layer.

\begin{figure}[tb!]
	\centering
	\begin{overpic}[width=0.7\textwidth]{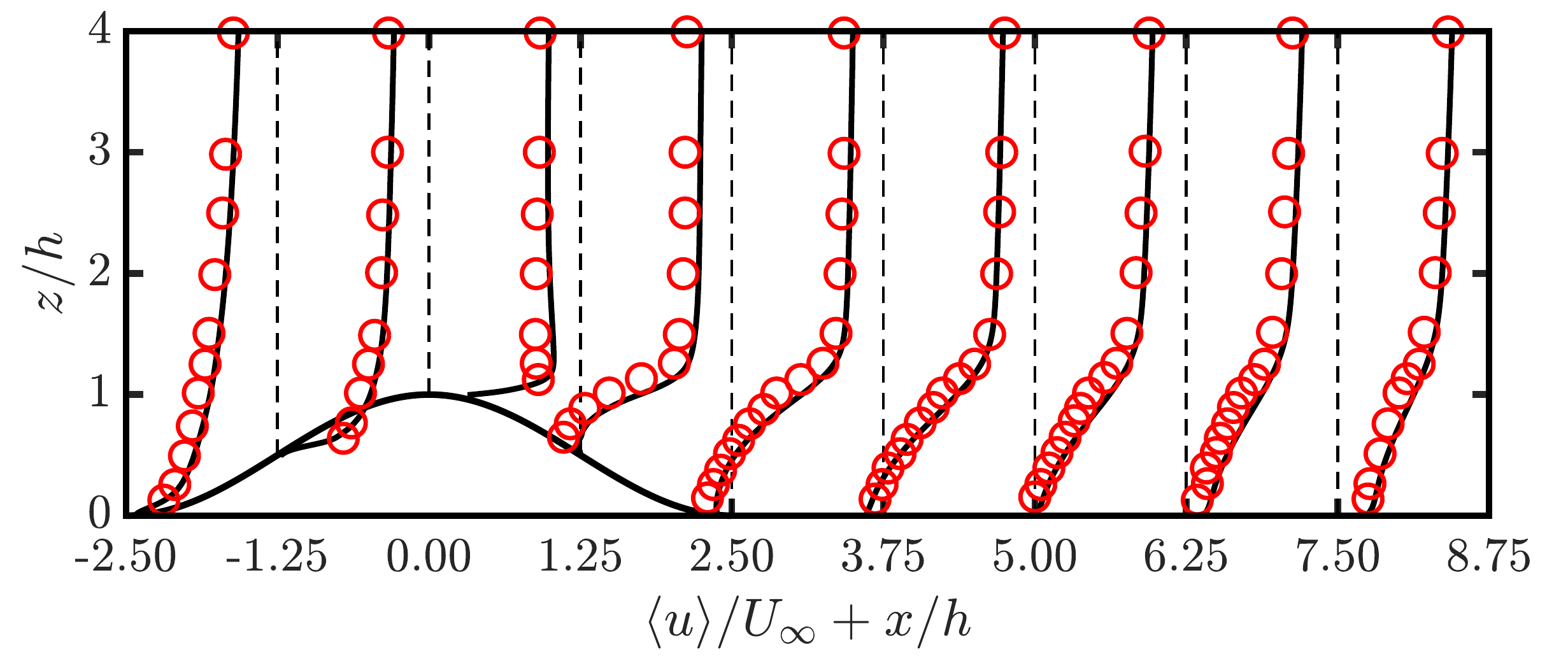}
		\put(0,37){\footnotesize $(a)$}
	\end{overpic}
	\begin{overpic}[width=0.7\textwidth]{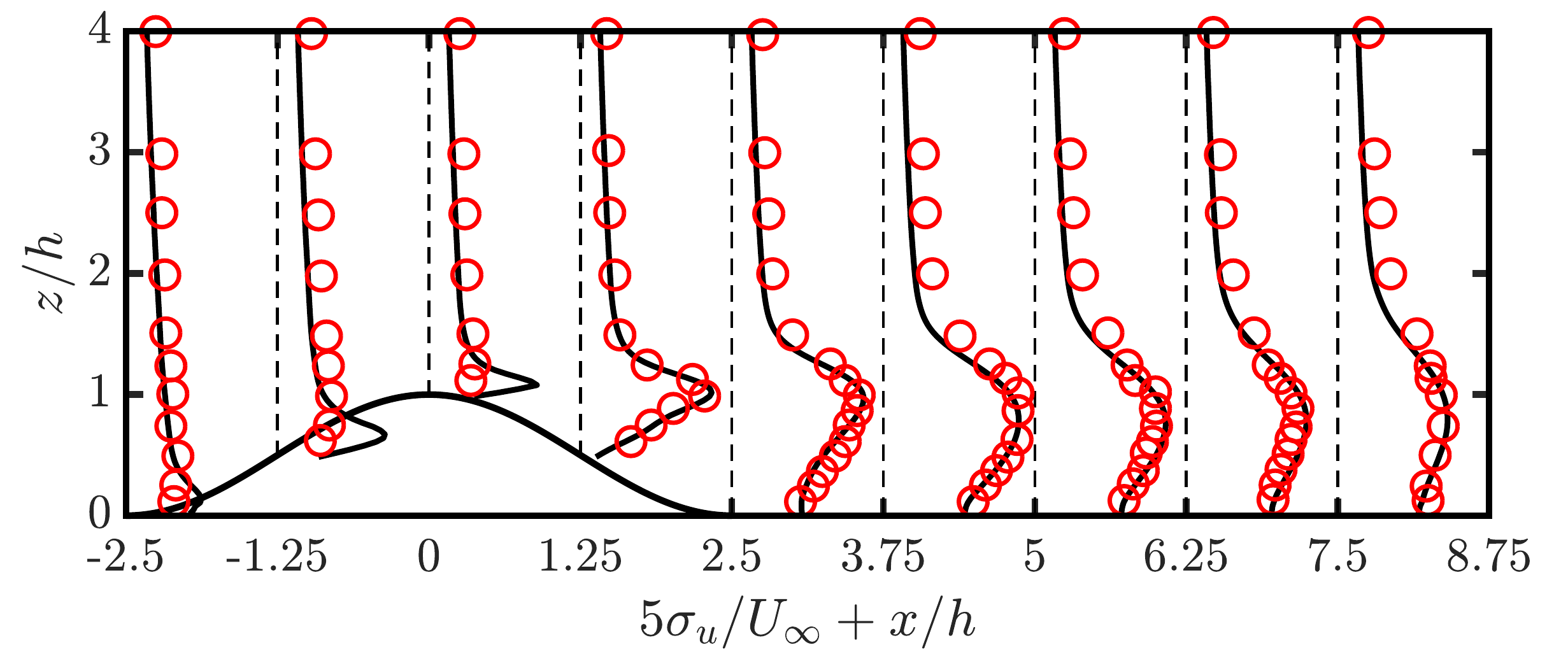}
		\put(0,37){\footnotesize $(b)$}
	\end{overpic}
	\caption{Horizontally and time averaged (a) streamwise velocity and (b) its variance profiles for atmospheric boundary layer over a two-dimensional ridge. Open circles: experimental data by \citet{cao06}, solid lines: simulation results. The vertical dashed lines indicate the measurement positions.}
	\label{fig.hill-u-2d}
\end{figure}

We use a simulation domain size $L_x \x L_y \x L_z = (5 \x 1 \x 1) L_z$, which is solved on a grid with $ 384 \x 128 \x 193$ nodes. Here, $L_z = 0.24$~m and the roughness height of the ridge surface is $z_{0, \rm IB} = z_0$. The center of the ridge is located at $x/L_x=1/4$. The inflow condition is generated with the concurrent precursor method \citep{ste14}. In this method, two domains are considered simultaneously. In the first domain, a neutral atmospheric boundary layer without a ridge is simulated to generate the turbulent inflow conditions for the second domain in which the ridge is placed. At the end of each time step a so-called fringe region is copied from the first domain to the end of the second domain to ensure a smooth transition from the flow formed behind the ridge towards the applied inflow condition. Figure~\ref{fig.hill-u-2d} shows that the vertical profiles of the streamwise velocity and its variance obtained by the simulations agree well with the wind tunnel data by \citet{cao06}. For instance, the relative difference of the streamwise velocity and its variance at $z-z_w=h$ obtained from the high resolution simulation and the experiments are less than 10\% and 20\%, respectively. The only significant differences are observed for the streamwise velocity variance at the second and third measurement stations. Close to the ground, the simulations indicate an increase in the velocity variance, which is not observed in the experimental data. We do not know the reason for this mismatch, but a similar behaviour has, for example, been observed in large-eddy simulations for turbulent flows over canopies \citep{pat06}.

\subsection{Complex terrain simulations: flow over three-dimensional hill} 
\lb{sec.hill-3d}

\begin{figure}[tb!]
	\centering
	\begin{overpic}[width=0.6\textwidth]{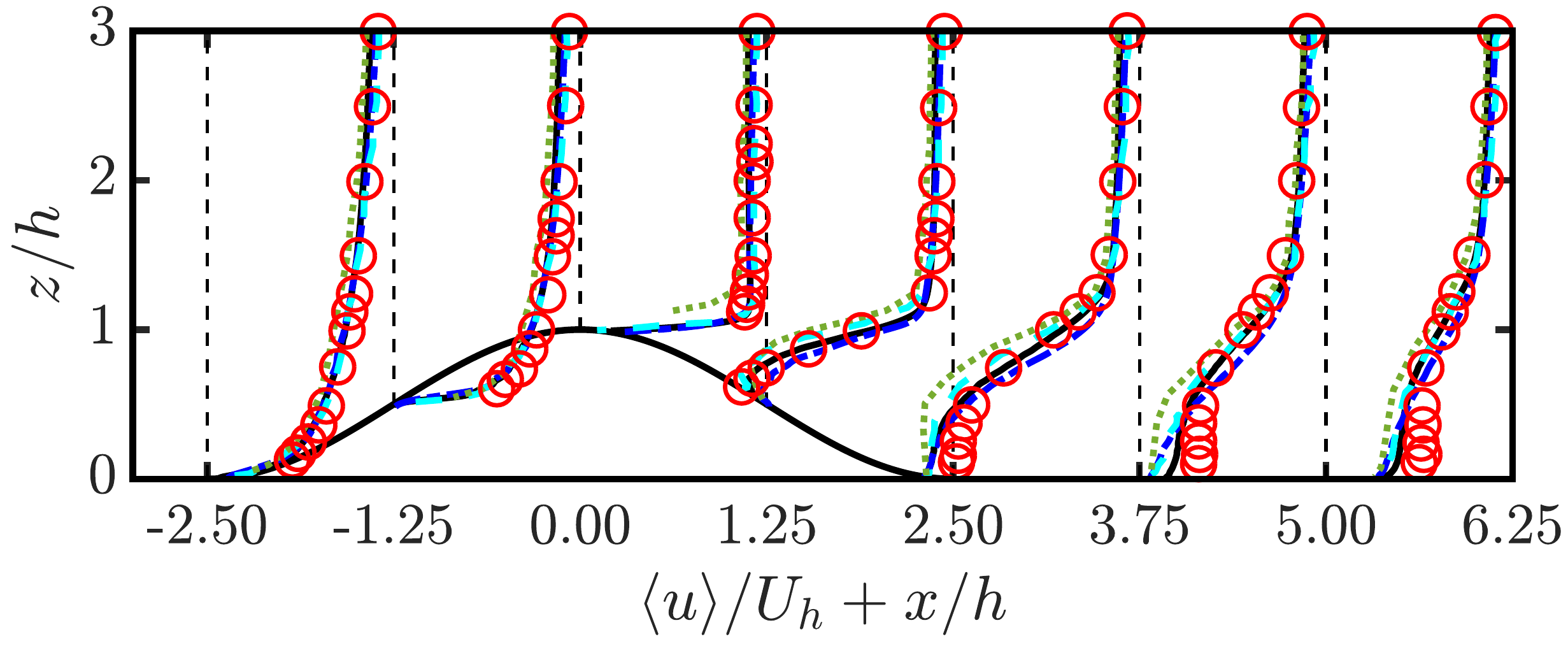}
		\put(0,36){\footnotesize $(a)$}
	\end{overpic}
	\begin{overpic}[width=0.6\textwidth]{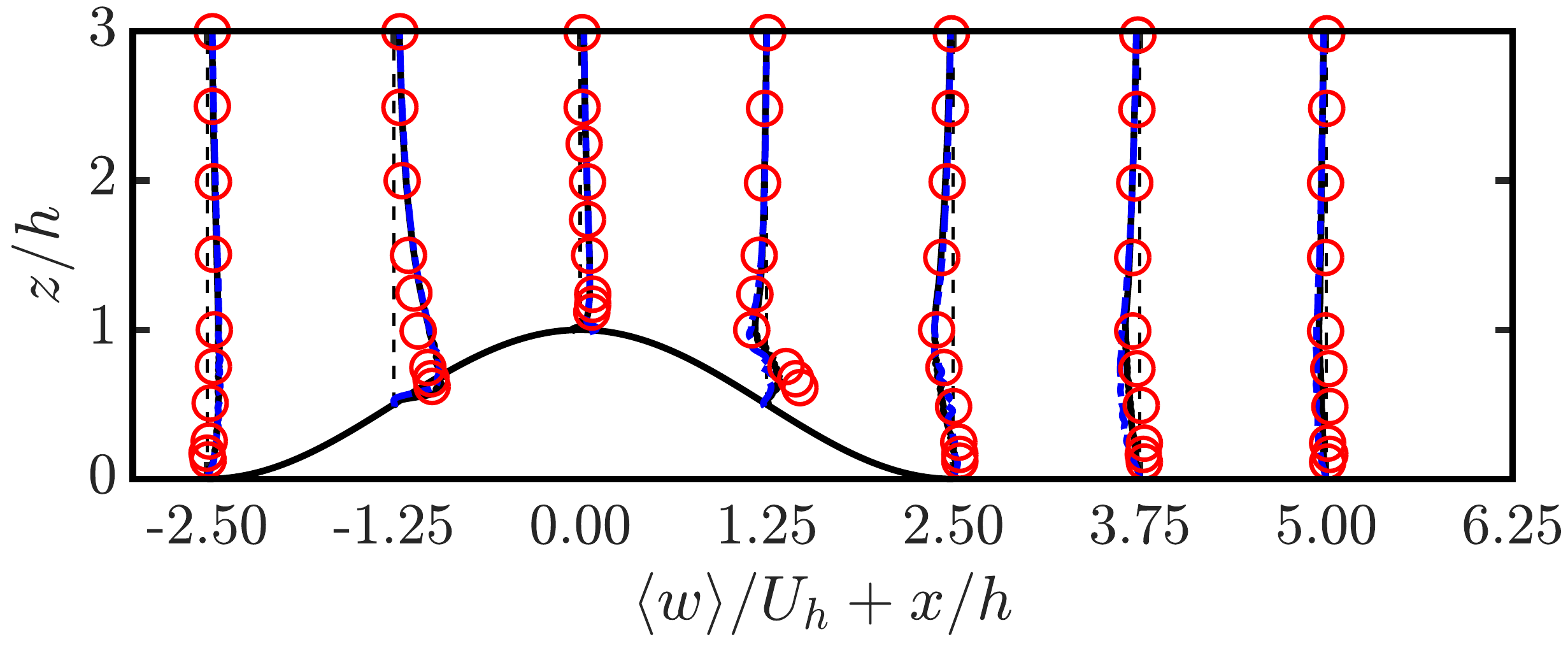}
		\put(0,36){\footnotesize $(b)$}
	\end{overpic}
	\caption{Time averaged (a) streamwise and (b) vertical velocity profiles over a three-dimensional hill. Open circles: experimental data by \citet{ish99}; green dotted lines: simulation results using the smearing method by \citet{die13}; cyan dashed lines: simulation results using the smearing method with normal derivative correction by \citet{fan16}; blue dashed-dotted lines: simulation results using the present method on a $384 \x 192 \x 192$ grid; black solid lines: simulation results using the present method on a $768 \x 384 \x 385$ grid. The vertical dashed lines indicate the measurement locations.}
	\label{fig.hill-u}
\end{figure}

\begin{figure}[tb!]
	\centering
	\begin{overpic}[width=0.6\textwidth]{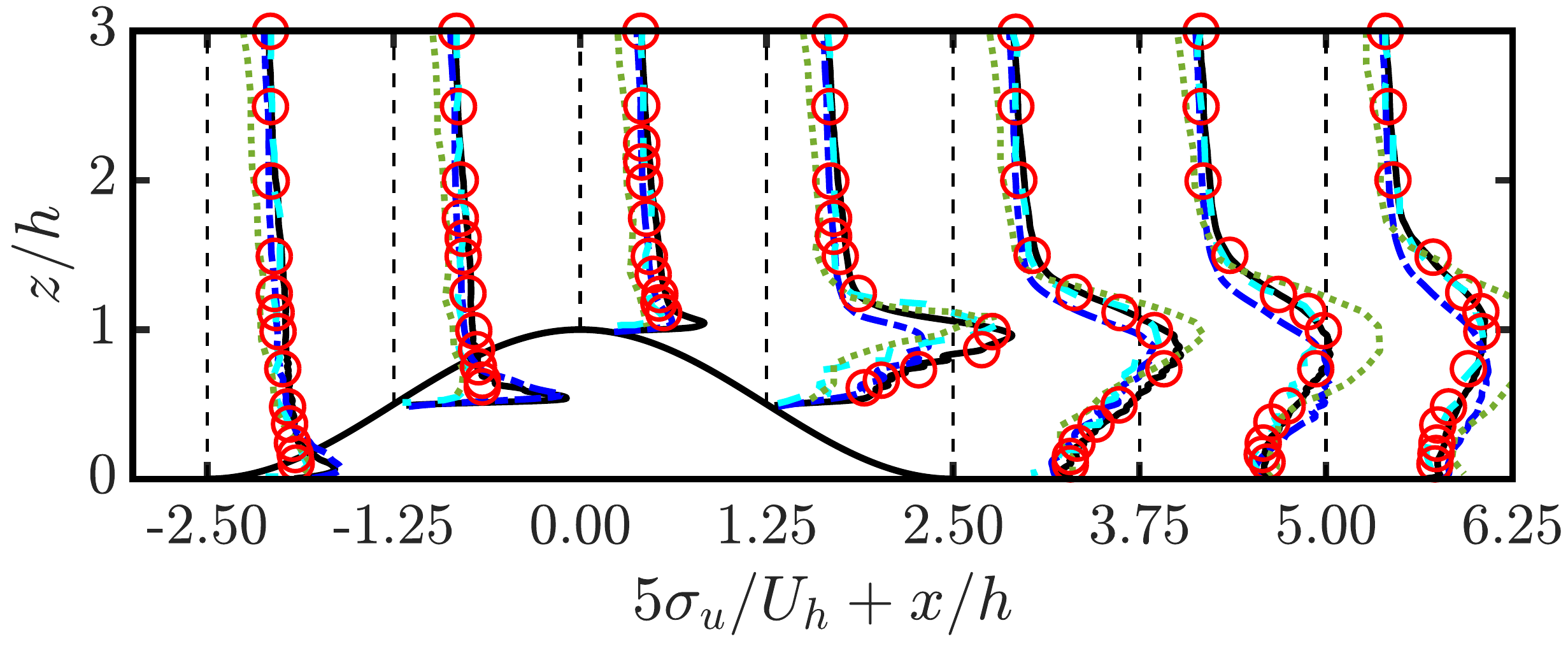}
		\put(0,36){\footnotesize $(a)$}
	\end{overpic}
	\begin{overpic}[width=0.6\textwidth]{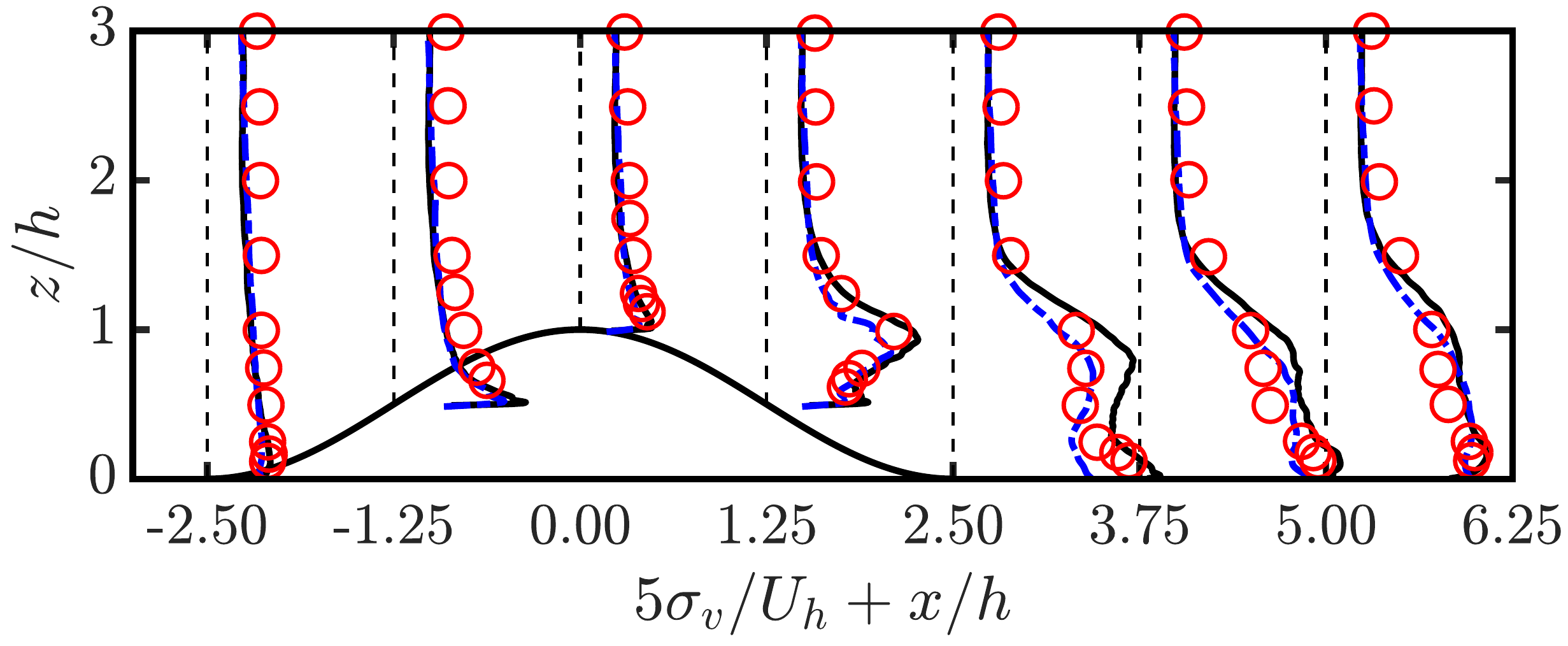}
		\put(0,36){\footnotesize $(b)$}
	\end{overpic}
	\begin{overpic}[width=0.6\textwidth]{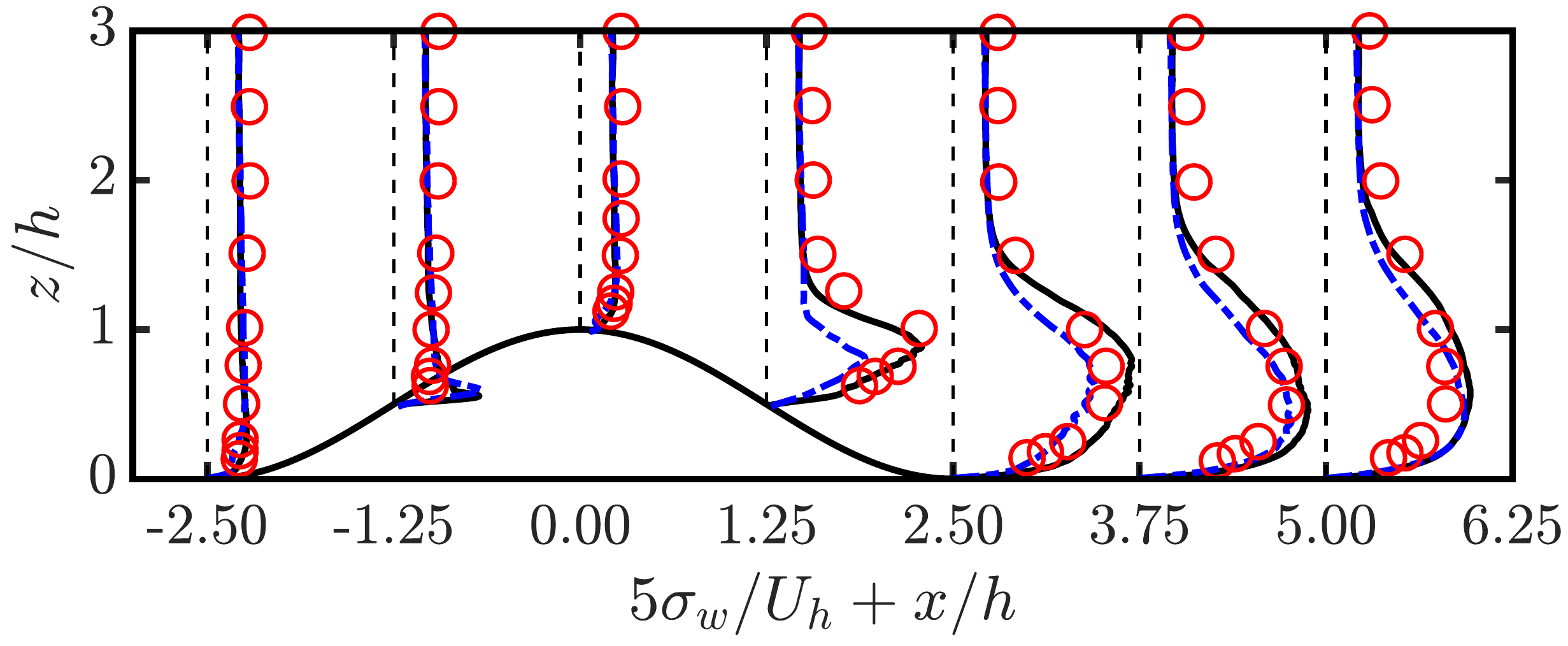}
		\put(0,36){\footnotesize $(c)$}
	\end{overpic}
	\caption{Time averaged variances profiles of the (a) streamwise, (b) spanwise, and (c) vertical velocity for flow over a three-dimensional hill. Open circles: experimental data by \citet{ish99}; green dotted lines: simulation results using the smearing method by \citet{die13}; cyan dashed lines: simulation results using the smearing method with normal derivative correction by \citet{fan16}; blue dashed-dotted lines: simulation results using the present method on a $384 \x 192 \x 192$ grid; black solid lines: simulation results using the present method on a $768 \x 384 \x 385$ grid. The vertical dashed lines indicate the measurement locations.}
	\label{fig.hill-urms}
\end{figure}

The last test case we consider is the flow over a three-dimensional cosine-squared hill. \citet{ish99} conducted a wind tunnel experiment of this case using a three-dimensional hill machined from wood. The height profile $z_w$ of the hill is well described by 
\beq\lb{eq.hill-3d}
z_w(x,y) = h \cos^2 \left( \fr{\pi \sqrt{x^2+y^2}}{2 b} \right), \quad \sqrt{x^2+y^2} \le b,
\eeq
where $h=0.04$~m and $b=0.1$~m are the height and half-width of the hill, respectively. The wind profiles were measured at seven different locations, i.e., $x/h=-2.5$, $-1.25$, 0, 1.25, 2.5, 3.75, 5, which are aligned with the central axis of the hill. Horizontal profiles were also performed at several locations above the hill and in its wake. The flow measured without the hill corresponded to a neutral atmospheric boundary layer with thickness $\delta = 0.36$~m. The estimated roughness of the plywood floor in the test section is $z_0 = 1.0 \times 10^{-5}$~m. The size of the simulation domain is $L_x \x L_y \x L_z = (4 \x 2 \x 1) L_z$ with $L_z = 0.32$~m, which is discretised on a grid with $384 \x 192 \x 193$ and $768 \x 384 \x 385$ nodes, respectively. The centre of the hill is located at $x/L_x=1/4$ and $y/L_y=1/2$. The roughness height of the hill surface is $z_{0, \rm IB} = 2 z_0$, see also \citet{fan16}. Figures~\ref{fig.hill-u} and \ref{fig.hill-urms} show that the vertical profiles of the velocity and its variance obtained at the different measurement locations are in very good agreement with the wind tunnel data \citep{ish99}. For example, the relative difference of the streamwise velocity and its variance at $z-z_w=h$ obtained from the high resolution simulation and the experiments are less than 5\% and 10\%, respectively. The velocity and its variance are normalized using $U_h$, i.e., the incoming mean streamwise velocity at $z=h$. Just as for the two-dimensional ridge, the simulations predict a somewhat higher velocity variance on the windward side of the hill.

\begin{figure}[tb!]
	\centering
	\begin{overpic}[width=0.35\textwidth]{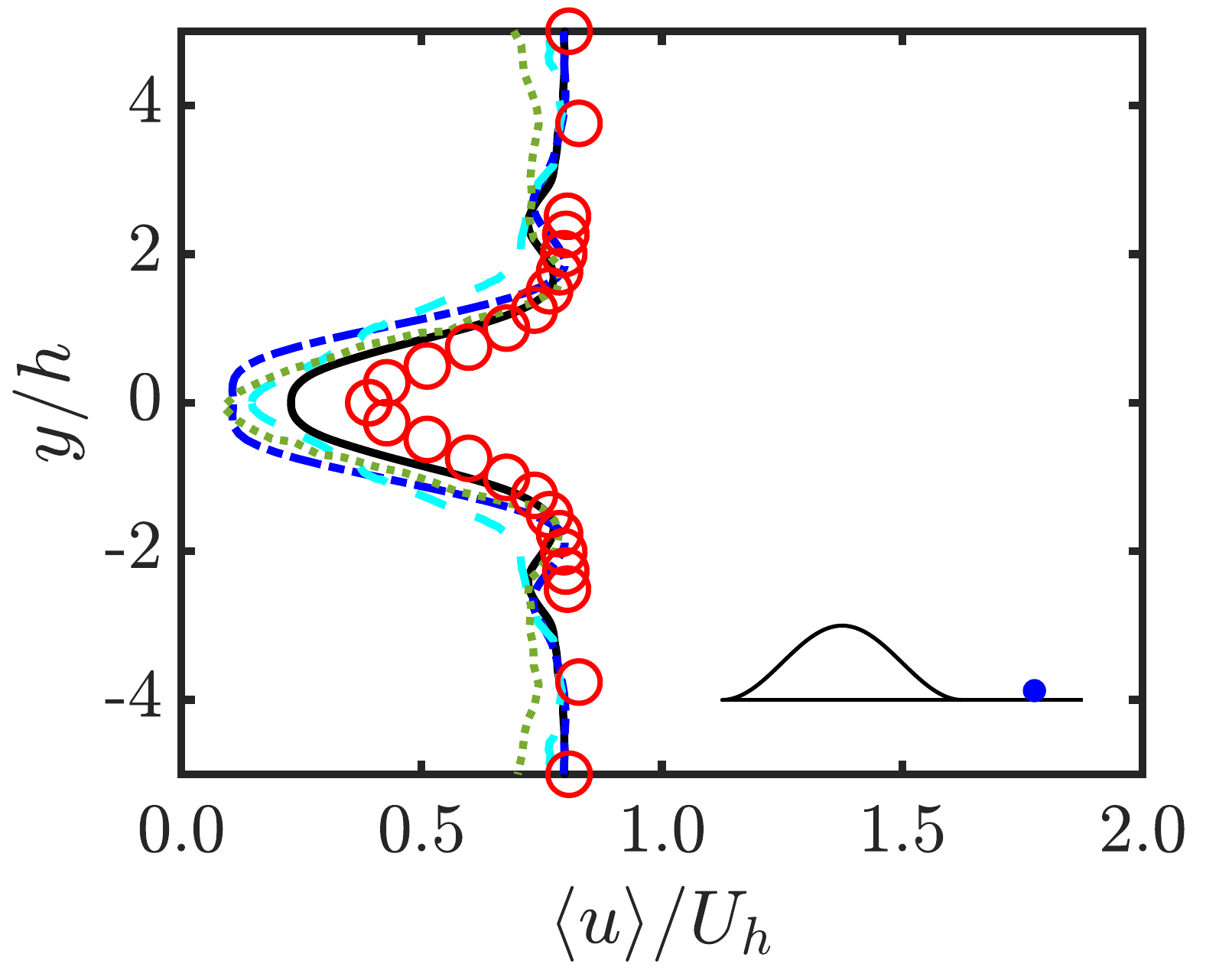}
		\put(0,74){\footnotesize $(a)$}
	\end{overpic}
	\begin{overpic}[width=0.35\textwidth]{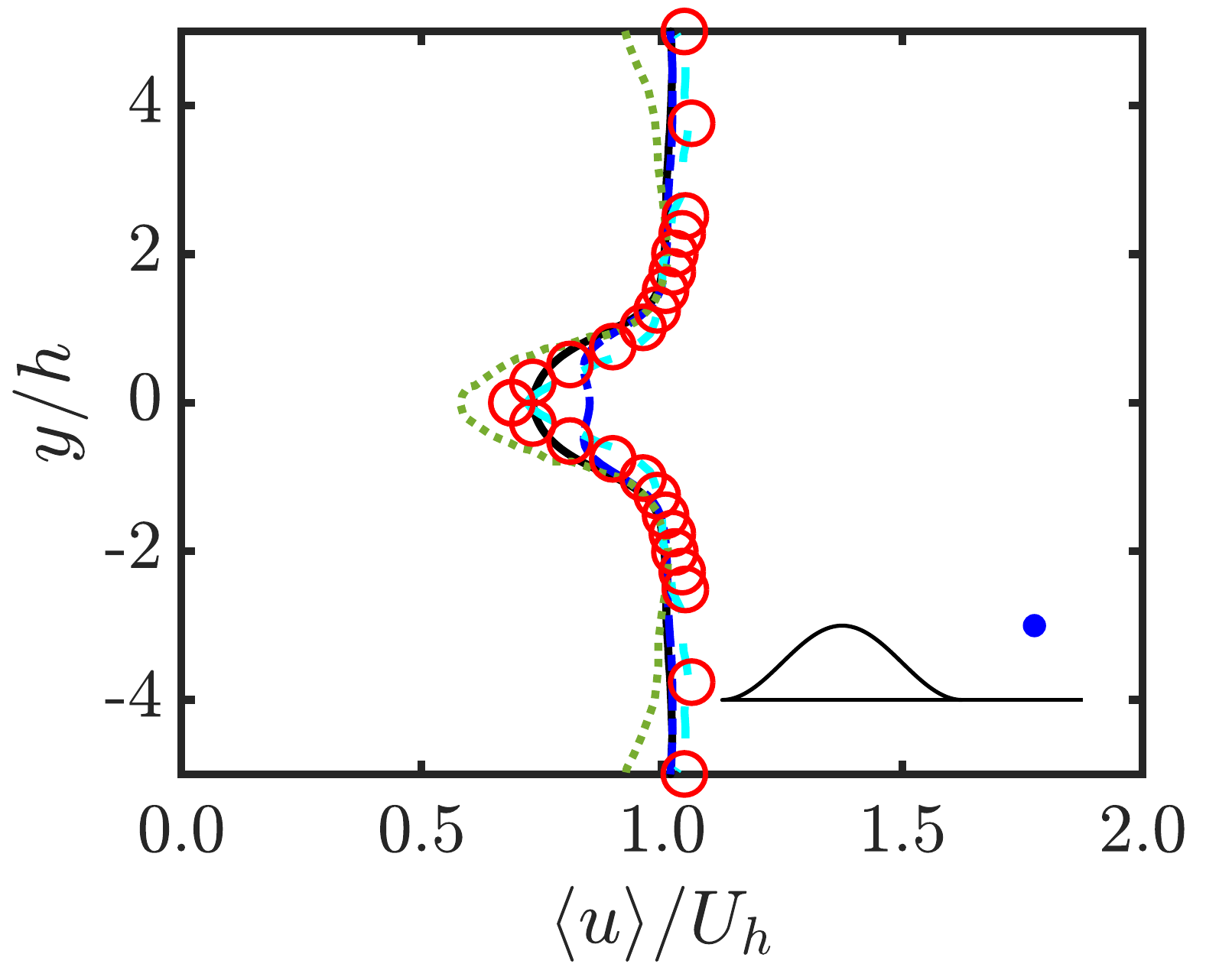}
		\put(0,74){\footnotesize $(b)$}
	\end{overpic} \\
	\begin{overpic}[width=0.35\textwidth]{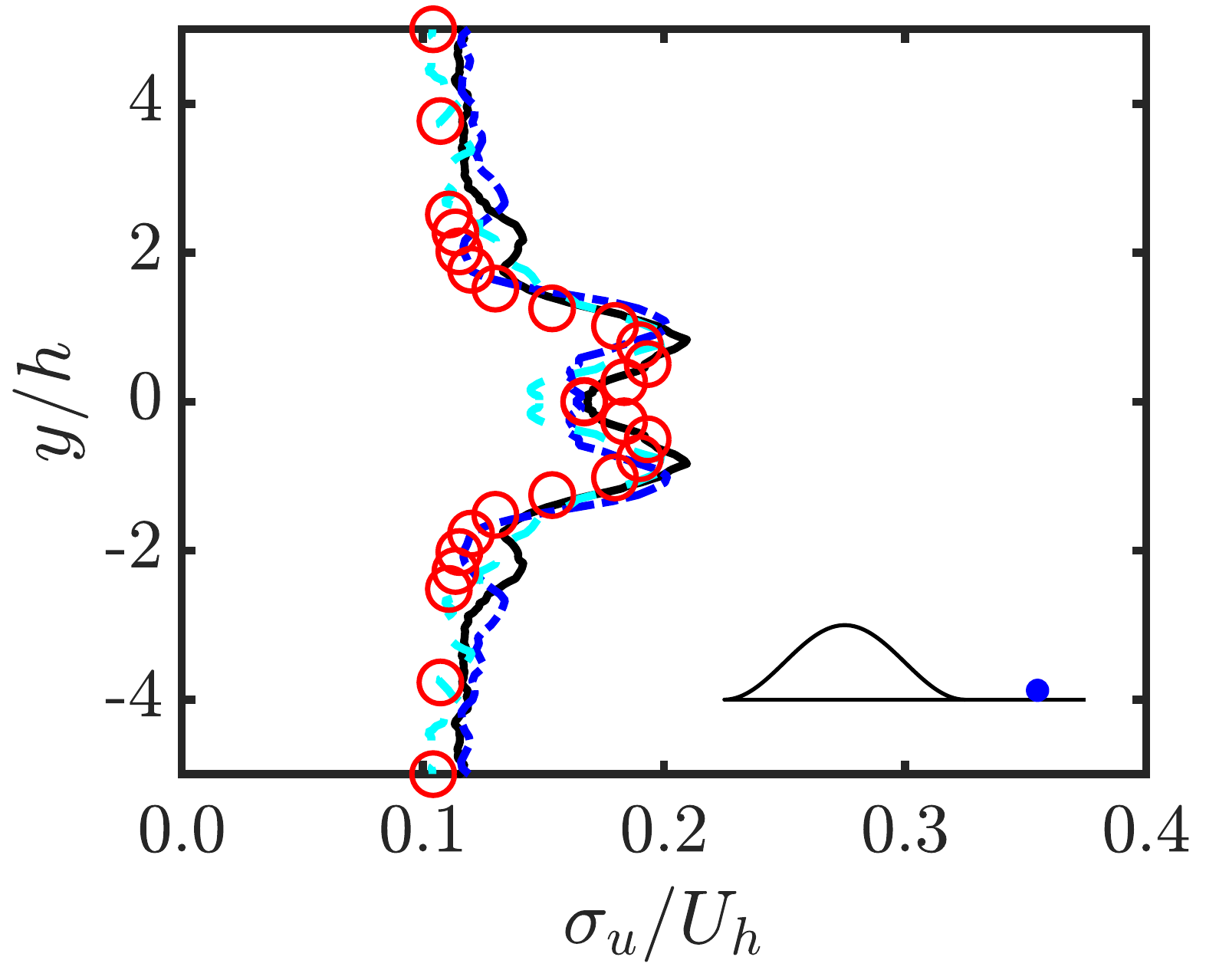}
		\put(0,74){\footnotesize $(c)$}
	\end{overpic}
	\begin{overpic}[width=0.35\textwidth]{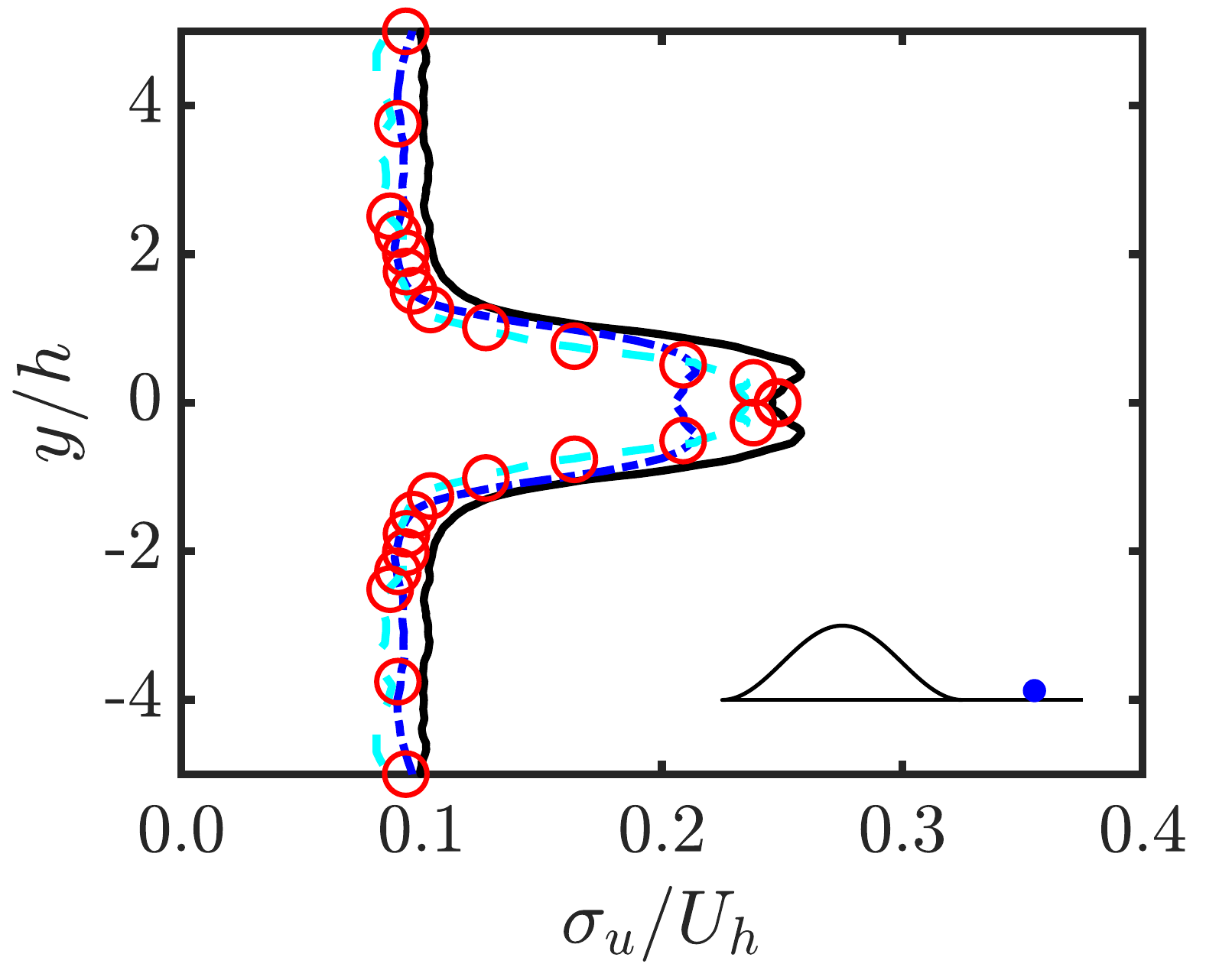}
		\put(0,74){\footnotesize $(d)$}
	\end{overpic}
	\caption{Time averaged (a, b) streamwise velocity and (c, d) the corresponding velocity variance at $x/h=3.75$. (a, c) $z/h=0.125$; (b, d) $z/h=1$. The blue dot in the sketch shown in the bottom right corners indicates the measurement location. Open circles: experimental data by \citet{ish99}; green dotted lines: simulation results using the smearing method by \citet{die13}; cyan dashed lines: simulation results using the smearing method with normal derivative correction by \citet{fan16}; blue dashed-dotted lines: simulation results using the present method on a $384 \x 192 \x 192$ grid; black solid lines: simulation results using the present method on a $768 \x 384 \x 385$ grid.}
	\label{fig.hill-u-23}
\end{figure}

It should be pointed out that most previous comparisons between large-eddy simulations and wind tunnel data focus exclusively on two-dimensional ridges and vertical profiles \citep{pat06, wan11, sha17}. The main advantage of the Ishihara dataset is that comparisons can also be made along the horizontal plane. Therefore, Fig.~\ref{fig.hill-u-23} shows a comparison of the simulation results with the experimental measurements at $x/h=3.75$ for two vertical locations ($z/h=0.125$ and $z/h=1$). The figure shows that the agreement at $z/h=1$ is also very good and that the maximum difference of the velocity and turbulence intensity obtained from the high resolution simulation and the experiments is only about 5\% and 15\%, respectively. We note that the simulations somewhat underestimate the velocity close to the ground ($z/h=0.125$) \citep{die13, fan16}. The reason is that the large-eddy simulation in which the wall effect is modeled is less accurate close to the wall. This is also the case when the wall modeled large-eddy simulation results for flow over a flat plate are compared to experimental measurements \citep{ste14d}. Besides, we note that some peaks in the turbulence intensity are captured less accurately when the numerical resolution is lower. Since the overall flow characteristics are captured reasonably accurately, we conclude that the wall modeled immersed boundary method is suitable to study the flow characteristics for flow over complex terrain. 

As remarked at the end of Section~\ref{sec.wall unsolved}, the proposed wall-modeled immersed boundary method is similar to the smearing method proposed by \citet{che07}. Therefore, while the results obtained using our method agree well with various experimental data sets it is still useful to see how it compares to previous simulation results for the \citet{ish99} case. Therefore, we compare our simulations with the results from \citet{die13}, who used the smearing method by \citet{che07} to simulate the three-dimensional hill case by \citet{ish99} on a $L_x \x L_y \x L_z = (2 \x 2 \x 1) L_z$ domain using a grid resolution of $256 \x 256 \x 129$, in Figs.~\ref{fig.hill-u}-\ref{fig.hill-u-23}. Aiming to improve the simulation accuracy of the smearing method, \citet{fan16} proposed a normal derivation correction for the velocity gradient near the solid surface and simulated the same hill case on a $L_x \x L_y \x L_z = (4 \x 2 \x 1) L_z$ domain using a grid resolution of $512 \x 256 \x 257$. The corresponding results are also shown in Figs.~\ref{fig.hill-u}-\ref{fig.hill-u-23}. The figures show that the results obtained using our proposed method compare quite favourably to the experimental measurement data. We note that the agreement of our results with the experimental data is similar to the agreement found by \citet{die13} and \citet{fan16}. However, as discussed at the end of Section~\ref{sec.flat}, our method offers various computational and implementation benefits compared to the original \citet{che07} method, which makes it an attractive approach for simulations performed on highly parallel computing platforms.

It is worth pointing out that we observe spurious oscillations in our simulations for the three-dimensional hill case when the numerical resolution is insufficient. These oscillations are due to the use of pseudo-spectral discretisation in the horizontal plane and are mainly created near the top of the hill, where sharp gradients in both horizontal directions are created. Therefore, this problem is more pronounced for the three-dimensional hill than for a two-dimensional ridge, which was discussed in Section~\ref{sec.hill-2d}. Here we limit the oscillations by using a sufficiently high grid resolution. In our simulations, we find that the required resolution to obtain good agreement with the experimental measurements for the velocity variances near the wall is sufficient to limit the effect of Gibbs oscillations. Motivated by this result, and given that we aimed to develop an efficient algorithm for parallel simulations, we decided not to include the smoothing technique here. The non-locality of smoothing operations, namely significantly reduces the computational efficiency of the proposed method.

We note that smoothing techniques are widely employed to reduce the spurious oscillations. In our proposed method, we set the velocity and the stress field inside the body to zero, except for the stresses at the grid points closest to the immersed boundary. We tried two different methods to reduce these spurious oscillations on coarser grids. The first method is to filter the velocity field or the stress field inside the body before performing the pseudo-spectral calculations \citep{tse06, che07, fan11, li16}, and the second one is to filter the Fourier coefficients after performing the pseudo-spectral calculations \citep{got97}. However, we found that neither method resulted in a significant improvement, which contributed to our decision to not include a smoothing method for now. This finding is consistent with the results of \citet{iac03}, who concluded that smoothing the internal velocity has virtually no effect on the oscillations when a direct forcing approach is used. Although, as argued above, we do not need the smoothing case for the test cases presented here, there could be cases in which the use of an additional smoothing method is beneficial.

\section{Conclusions}\lb{sec.conclusion}

In this paper, we presented and validated a simplified version of the wall modeled immersed boundary method by \citet{che07} for large-eddy simulations of high Reynolds number flows. Considering the trend that simulations are performed using an ever-increasing number of cores, we believe that it is essential to have a simpler more computational efficient method that is very efficient in highly parallel computations. Particular attention has been given to an efficient reconstruction of the wall stresses near the immersed boundary to ensure that the vertical stress-balance, which is important for turbulent boundary layer flow, is maintained. The method is developed for wall modeled high Reynolds number flows in which the viscous sublayer is not explicitly resolved. This means that accurate interpolation of velocities in the near-wall region is impossible, and based on this notion, some approximations are made to ensure the vertical stress-balance in the turbulent boundary layer is enforced. This is achieved by determining the wall-normal stresses at the grid locations closest to the immersed boundary where the corresponding wall-normal velocity should be zero. As a result, the other stress components have to be neglected to make sure that the vertical stress-balance is maintained. An efficient implementation is obtained by identifying and characterizing the grid points for which the wall-stresses are modeled when the simulation is initialized. A benefit of the chosen approach is that it limits the communication overhead between different processors. The suitability of the proposed method for parallel computations is underlined by the low computational overhead, which is less than $2\%$. The method is also attractive as it is relatively easy to implement.

We find that the results from the wall modeled immersed boundary layer method agree very well with the results from a reference large-eddy simulation when the immersed boundary coincides with the grid locations. However, some unavoidable and relatively small effects of the introduced approximations are observed when the immersed boundary does not coincide with the grid locations. The effect of these approximations is assessed by comparing simulation results for cases in which the immersed boundary does not coincide with the computational grid with the reference solution. In addition, we validate our approach against wind tunnel measurements for flow over wall-mounted cubes, a two-dimensional ridge, and a three-dimensional hill. In all these complex terrain simulations, both the vertical and horizontal profiles of the mean velocities and the Reynolds stresses are in good agreement with the corresponding experimental measurements. We find, in agreement with the results of \citet{iac03}, that smoothing of the velocity field in the body has virtually no effect on the spurious oscillations. However, we find that for the three-dimensional hill case numerical oscillations are not triggered anymore when a sufficiently high resolution, which is anyhow required to resolve important flow features, is used. Overall, the good agreement of the simulation results with the experimental data shows that our straightforward and computationally efficient method is an attractive approach to consider wall modeled immersed boundary method in large eddy simulations of high Reynolds number flows. 

\section*{Acknowledgments}
\addcontentsline{toc}{section}{Acknowledgements}

We appreciate insightful suggestions by anonymous referees that have helped us to improve the manuscript. We thank Srinidhi Nagarada Gadde, Anja Stieren, and Jessica Strickland for discussions, and Jason Graham for providing the experimental data by \citet{mei99} for the wall mounted cubes case. This work is part of the Shell-NWO/FOM-initiative Computational sciences for energy research of Shell and Chemical Sciences, Earth and Live Sciences, Physical Sciences, FOM and STW and an STW VIDI grant (No.\ 14868).\ This work was carried out on the national e-infrastructure of SURFsara, a subsidiary of SURF cooperation, the collaborative ICT organization for Dutch education and research.

\addcontentsline{toc}{section}{References}
\bibliographystyle{model3-num-names}
\bibliography{literature_windfarms}

\end{document}